\documentclass[%
 aip,
pop,
 amsmath,amssymb,
preprint,%
]{revtex4-1}

\usepackage{graphicx}
\usepackage{dcolumn}
\usepackage{bm}
\usepackage[utf8]{inputenc}
\usepackage[T1]{fontenc}
\usepackage{etoolbox}
\usepackage{physics}

\makeatletter
\def\@email#1#2{%
 \endgroup
 \patchcmd{\titleblock@produce}
  {\frontmatter@RRAPformat}
  {\frontmatter@RRAPformat{\produce@RRAP{*#1\href{mailto:#2}{#2}}}\frontmatter@RRAPformat}
  {}{}
}%
\makeatother
\begin{document}
\title{Direct comparison of the energization of self-consistent charged particles vs test particles in a turbulent plasma}

\author{F.Pugliese}
\affiliation{CONICET - Universidad de Buenos Aires, Instituto de Física Interdisciplinaria y Aplicada (INFINA), Ciudad Universitaria, 1428 Buenos Aires, Argentina.}
\affiliation{Universidad de Buenos Aires, Facultad de Ciencias Exactas y Naturales, Departamento de Física, Ciudad Universitaria, 1428 Buenos Aires, Argentina.}

\author{P. Dmitruk}
\affiliation{CONICET - Universidad de Buenos Aires, Instituto de Física Interdisciplinaria y Aplicada (INFINA), Ciudad Universitaria, 1428 Buenos Aires, Argentina.}
\affiliation{Universidad de Buenos Aires, Facultad de Ciencias Exactas y Naturales, Departamento de Física, Ciudad Universitaria, 1428 Buenos Aires, Argentina.}

\begin{abstract}
The test particle approach is a widely used method for studying the dynamics of charged particles in complex electromagnetic fields and has been successful in explaining particle energization in turbulent plasmas. However, this approach is fundamentally not self-consistent, as test particles do not generate their own electromagnetic fields and therefore do not interact with their surroundings realistically. In this work, we compare the energization of a population of test protons in a magnetofluid to that of a plasma composed of self-consistent particles. We use a compressible Hall magnetohydrodynamic (CHMHD) model for the test particle case and a hybrid particle-in-cell (HPIC) approach for the self-consistent case, conducting both 2D and 3D simulations. We calculate the rate of energization and conversion to thermal energy in both models, finding a higher temperature for the test particle case. Additionally, we examine the distribution of suprathermal particles and find that, in the test particle scenario, these particles eventually occupy the entire domain, while in the self-consistent case, suprathermal particles are confined to specific regions. We conclude that while test particles capture some qualitative features of their self-consistent counterparts, they miss finer phenomena and tend to overestimate energization.
\end{abstract}

\maketitle

\section{Introduction} \label{sec:intro}

Solar energetic particles (SEP) are high energy charged particles generated by the sun that propagate through the interplanetary medium.
Some of these SEP reach Earth and contribute to the cosmic radiation constantly bombarding the planet.
Multiple solar processes contribute to the generation of this particular form of cosmic rays, such as solar flares\cite{Lin1971, Reames1994, Miller1997} or coronal mass ejections \cite{Desai2016, Vlahos2019}.
Others are produced within the solar wind, a constant flux of charged particles emanating from the sun\cite{Parker1958}.
The production of such high energy particles is one of the proposed mechanisms to explain the non adiabatic expansion of the solar wind\cite{Matthaeus1999, Hellinger2013}.
This heating process should be collisionless in origin, as the mean free paths of charged particles within the solar wind is rather large\cite{Verscharen2019}.
Therefore, electromagnetic fields with fluctuations comparable to particle scales are needed and the prime candidate mechanism for generating them is turbulence\cite{Matthaeus2011, Bruno2013}.
Direct energy cascades in which energy flows from large to progressively smaller scales are a characteristic trait in turbulent systems\cite{Frisch1995} and has been observed within the solar wind\cite{SorrisoValvo2007, Hadid2017, Andres2019, Romanelli2022}.
Once this energy reaches ion or electron scales, particles are able to exploit coherent structures in the electromagnetic fields to achieve high kinetic energy.

Particle interactions with such structures are very complex and there is yet no definitive model to predict energization rates\cite{Stix1992, Bieber2004, Ruffolo2008}.
Within those models, the use of test particles is a popular and rather cheap approach for computing energization, where particles do not generate electromagnetic fields but merely react to externally imposed ones.
As such, the validity of their predictions is mostly determined by how realistically modeled are the external electromagnetic fields, either by synthetic turbulence\cite{Ruffolo2006, Minnie2007, Dalena2012, Tautz2013, Dalena2014} or that obtained by direct numerical simulation\cite{Dmitruk2004, dmitruk2006b, Gonzalez2016, Gonzalez_2017, pugliese2022, Pezzi2022, Balzarini2022}.
While the former is computationally cheap and able to capture global statistics such as diffusion coefficients, the latter is more suited to reproduce commonly known structures presents in the plasma, which can be very relevant in particle energization\cite{Greco2014, Lemoine2021, Pezzi2022, pugliese2022, pugliese2023}.
Test particles are most commonly used to study high energy but low concentration charged particle dynamics, lest their effect on the external fields becomes noticeable and their intrinsic electromagnetic fields must be taken into account.

Such high energy particles are not the only product of collisionless heating in a plasma, but also other suprathermal particles with considerably less energy and higher concentration.
In order to capture this production, self-consistent kinetic models that include the particle feedback on the fields are required\cite{Servidio2017, Howes2017}.
Fully kinetic models solve both ion and electron dynamics, with the latter having considerably smaller characteristic scales and thus demanding higher spatial and temporal resolutions.
When interested mainly on ions, this increased resolution leads to an increase in computational cost which can limit the simulation size without providing more interesting (ion) physics.
To counter this, hybrid models were developed in which only ions are treated kinetically while electron are treated as a fluid, usually massless (see, however, Muñoz 2018 and references therein for hybrid models that include electron mass\cite{Muñoz2018}).
Not only do these models greatly reduce computational cost, but also show good agreement with fully kinetic simulations in reproducing ion-scale phenomena, including wave-particle interactions, shock dynamics, and energy transfer processes\cite{Birn2001,Le2016,Gonzlez2023}.

The main objective of the present work is to assess to which extent the test particle approximation can reproduce the production of suprathermal ions (protons) on turbulent plasmas in the presence of magnetic guide field.
The externally supplied electromagnetic fields will be those obtained from magnetohydrodynamics, with the inclusion of the Hall term to capture (massless) electron effects at sub-ionic scales.
As a ground truth for comparison, we use a self-consistent formulation based on a hybrid kinetic model, which we consider more realistic as previously discussed.
While it is possible to add different kinetic ion species, here we are interested in a plasma composed only of protons and electrons, as they are the most prominent species in the solar wind.

This paper is organized as follows: in section \ref{sec:theo}, we introduce the theoretical set of equations of the hybrid kinetic and the compressible Hall magnetohydrodynamics (CHMHD) models, including the relevant parameters that enable comparison between the two.
In section \ref{sec:num_set}, we describe the numerical methods used to solve such equations, with special emphasis on the hybrid particle-in-cell (HPIC) method.
We also explain how the initial conditions for both models are constructed and how test and self-consistent particles are treated.
Finally, in section \ref{sec:res} we present our results and in section \ref{sec:disc} we discuss our main findings and their implications.

\section{Theory}\label{sec:theo}
\subsection{Hybrid model equations}\label{ssec:HPICeq}

In the hybrid model, ions are treated as particles following collisionless kinetic equations while electrons are considered a massless fluid.
Therefore, the particle distribution function $f_i(\vb{x},\vb{v},t)$ in phase space for ions follows the classical collisionless Vlasov equation
\begin{equation}
    \pdv{f_i}{t} + \vb{v}\cdot\pdv{f_i}{\vb{x}} + {q_i \over m_i}\left(\vb{E} + {\vb{v}\over c}\cross \vb{B}\right)\cdot\pdv{f_i}{\vb{v}} = 0, \label{eq:vlasov_ions}
\end{equation}
where $q_i$ and $m_i$ are the ion electric charge and mass, respectively.
The electric field $\vb{E}$ and the magnetic field $\vb{B}$ are prescribed by the generalized Ohm's law and Maxwell equations.
As usual, Ohm's law is obtained from the momentum equation of the electron fluid by setting the electron mass $m_e=0$,
\begin{equation}
    \vb{E} = -{\vb{u}_e \over c}\cross\vb{B} - {\grad p_e \over en_e}, \label{eq:gen_Ohm}
\end{equation}
where $n_e$ is the electron particle density, $\vb{u}_e$ the electron bulk velocity field, $p_e$ the isotropic electronic pressure and $e$ is the electron charge (in modulus).
To obtain these electronic fields, we apply a series of additional hypothesis.
For the electron particle density $n_e$ we assume quasi-neutrality, thus setting the charge density to zero.
This assumption yields
\begin{equation}
    n_e = {q_i \over e} n_i, \label{eq:quasineutrality}
\end{equation}
where $n_i$ is the ion particle density, obtained as the zeroth moment in velocity space of the ion particle distribution function $f_i$
\begin{equation}
    n_i(\vb{x}, t) = \int f_i(\vb{x}, \vb{v}, t) \dd \vb{v}, \label{eq:def_ni}
\end{equation}

For the electron bulk velocity $\vb{u}_e$, we use the Ampere-Maxwell law without the radiative term (quasi-stationary approximation)
\begin{equation}
    \curl\vb{B} = \frac{4\pi}{c}\left(q_in_i\vb{u}_i - en_e\vb{u}_e\right) \equiv \frac{4\pi}{c}\vb{j},\label{eq:ampere_law}
\end{equation}
where $\vb{j}$ is the current density and $\vb{u}_i$ is the ion bulk velocity, obtained as the first moment in velocity space of the ion particle distribution function $f_i$
\begin{equation}
    n_i(\vb{x}, t)\vb{u}_i(\vb{x}, t) = \int \vb{v}f_i(\vb{x}, \vb{v}, t) \dd \vb{v}. \label{eq:def_ui}
\end{equation}

Combining Ampere's law \eqref{eq:ampere_law} with quasineutrality \eqref{eq:quasineutrality} we obtain an expression for $\vb{u}_e$ in terms of the magnetic field $\vb{B}$ and ionic fields
\begin{equation}
    \vb{u}_e = \vb{u}_i - {c \over 4\pi e n_i}\curl\vb{B},
\end{equation}

Finally, the electronic pressure $p_e$ is obtained in terms of $n_e$ assuming an adiabatic relation
\begin{equation}
    p_e = p_{e,0}\left({n_e \over n_{0}}\right)^{\gamma}, \label{eq:adiabatic_electrons}
\end{equation}
where $p_{e,0}$ and $n_{0}$ are some reference pressure and particle density, respectively, and $\gamma=5/3$ is the adiabatic exponent for a monoatomic ideal gas.
From these reference values it is possible to obtain a reference electronic temperature $T_{e,0} = p_{e,0}/n_0$.

In order to completely close the system of equations, we use Faraday's Law to determine the evolution of $\vb{B}$
\begin{equation}
    \pdv{\vb{B}}{t} = -c\curl\vb{E} = \curl\left( \vb{u}_i \cross\vb{B} \right) - {c \over 4\pi e}\curl\left[ {\curl\vb{B} \over n_i}  \cross\vb{B} \right], \label{eq:faraday_law_hyb}
\end{equation}
where we made use of the fact that $\nabla p_e/n_e$ can be written as a gradient thanks to the adiabatic relation \eqref{eq:adiabatic_electrons} and therefore has zero curl.

In general, we are interested in the case where a constant magnetic guide field $\vb{B_0}=\abs{\vb{B_0}}\vu{z}$ is present and therefore we decompose the total magnetic field as $\vb{B} = \vb{B_0} + \vb{b}$ where $\vb{b}$ is the fluctuating part with zero mean.
In other words, $\langle \vb{b} \rangle = 0$ where we denote $\langle \bullet\rangle$ as the average over the whole domain.
Therefore, Eq.~\eqref{eq:faraday_law_hyb} is actually an evolution equation for $\vb{b}$.

From these system of equations it is possible to derive a set of conserved quantities, which we identify as energy densities but we will just call energies from now on.
The first of these densities is the ion kinetic energy $\mathcal{E}_i$, which can be split in a bulk kinetic energy $\mathcal{E}_b$ plus a fluctuation kinetic energy (which we identify as thermal) $\mathcal{E}_{\mathrm{th}}$ such that
\begin{equation}
    \mathcal{E}_b(\vb{x},t) = \frac{1}{2}m_in_i \abs{\vb{u}_i(\vb{x},t)}^2, \label{eq:def_Eb}
\end{equation}
\begin{equation}
    \mathcal{E}_{\mathrm{th}}(\vb{x},t) = \frac{1}{2}m_i\int \abs{\vb{v}-\vb{u}_i(\vb{x})}^2f_i(\vb{x},\vb{v},t)\dd\vb{v} \equiv \frac{3}{2}n_i(\vb{x},t) T_i(\vb{x},t), \label{eq:def_Eth}
\end{equation}
where we have also defined the ion temperature $T_i$.
Furthermore, there is a thermal electronic energy
\begin{equation}
    \mathcal{E}_e(\vb{x},t) = {p_e(\vb{x},t)\over \gamma - 1}. \label{eq:def_Ee}
\end{equation}

Finally, the magnetic energy $\abs{\vb{B}}^2/8\pi$ can be decomposed as the magnetic field itself in a constant part $\abs{\vb{B_0}}^2/8\pi$ and a fluctuating part
\begin{equation}
    \mathcal{E}_m(\vb{x},t) = \frac{1}{8\pi}\abs{\vb{b}(\vb{x},t)}^2, \label{eq:def_Em}
\end{equation}
which is the one we will be interested in.
The total mean energy $\langle\mathcal{E}\rangle = \langle\mathcal{E}_b + \mathcal{E}_{\mathrm{th}} + \mathcal{E}_e + \mathcal{E}_m\rangle$ is therefore a conserved quantity in absence of injection or dissipation.

\subsection{Compressible Hall MHD equations}\label{ssec:CHMHDeq}

The compressible Hall magnetohydrodynamic (CHMHD) model is obtained by taking the model in Sec. \ref{ssec:HPICeq} and also treating ions as a fluid.
This implies taking the zeroth and first order moments in Eq.~\eqref{eq:vlasov_ions} in combination with Eq.~\eqref{eq:gen_Ohm} to produce equations for the evolution of $\rho \equiv m_in_i$ and $\vb{u}_i$ (from now on, simply $\vb{u}$). 
It is customary to add a viscous term in the evolution equation of $\vb{u}$, to account for ion collisions and regularize solutions, yielding
\begin{equation}
    \frac{\partial \rho}{\partial t}+\boldsymbol\nabla\cdot(\rho \mathbf{u})=0,
    \label{eq:continuity}
\end{equation}
\begin{equation}
     \rho\left[\pdv{\vb{u}}{t}+(\vb{u}\cdot\grad) \vb{u} \right] =-\grad p +\frac{(\curl\vb{B})\cross\mathbf{B}}{4\pi} + \mu\left[\laplacian\vb{u} +\frac{\grad(\grad\cdot\vb{u})}{3}\right],
     \label{eq:NS}
\end{equation}
where $\mu$ is the dynamic viscosity coefficient, $p=p_i+p_e$ is the total pressure and, $p_i \equiv n_i T_i$ is the ionic pressure.
As we are interested in the solar wind, we will take the ions to be protons, thus setting $q_i=e$.
Therefore, we have $n_i=n_e$ and by further assuming thermodynamical equilibrium between both species $T_i=T_e$, pressures must be equal $p_i=p_e$ and Eq.~\eqref{eq:adiabatic_electrons} implies
\begin{equation}{\label{eq:adiabatic}}
    p = p_0\left({\rho \over \rho_0}\right)^\gamma,
\end{equation}
where $\rho_0=m_in_0$ is a reference mass density and $p_0=2p_{e,0}$.
This reduced system retains both bulk kinetic and magnetic energies, as defined in Eqs.~\eqref{eq:def_Eb} and \eqref{eq:def_Em}, but ion and electron thermal energies merge into a single thermal energy
\begin{equation}
    \mathcal{E}_{\mathrm{th}+e} = {p \over \gamma - 1}, \label{eq:def_Eth+e}
\end{equation}
which is indeed the sum of $\mathcal{E}_{\mathrm{th}}$ and $\mathcal{E}_e$, using that $\gamma = 5/3$ and $p_i = n_i T_i$.
Therefore, in the CHMHD model, we cannot separate the ionic and electronic components of the thermal energy, as we have assumed them to be equal.
The magnetic field $\vb{B}$ is treated similarly to the hybrid model, with an added resistivity $\eta$ to Eq.~\eqref{eq:faraday_law_hyb}
\begin{equation}
    \pdv{\vb{B}}{t} = \curl\left( \vb{u} \cross\vb{B} \right) - {m_ic\over 4\pi e }\curl\left[ {\curl\vb{B}\over\rho} \cross\vb{B} \right] + \eta\laplacian\vb{B}. \label{eq:faraday_law_chmhd}
\end{equation}

\subsection{Parameters and characteristic scales}\label{ssec:params_def}

The previous formulations allow for a rather straightforward comparison of the fields and parameters in both models.
For particle density and mass density, we simply use $n_0$ and $\rho_0=m_in_0$.
The existence of a guide field $\vb{B_0}$ allows us to define a characteristic time in terms of an ion gyrofrequency $\Omega_i = q_i\abs{\vb{B_0}}/m_ic$, with an associated gyroperiod $2\pi/\Omega_i$.
Furthermore, we can define a characteristic velocity $v_A = \abs{\vb{B_0}}/\sqrt{4\pi \rho_0}$, the velocity at which Alfvén waves propagate.
These choices yield a characteristic length $v_A/\Omega_i = c\sqrt{m_i/4\pi n_0 e} \equiv d_i$ known as the ion inertial length, which is independent of the guide field.
This ion inertial length $d_i$ coincides with the so called Hall scale, below which whistler waves become relevant and ion kinetic effects may appear.
It can be easily be related to the factor $m_ic/4\pi e$ in Eq. \eqref{eq:faraday_law_chmhd}.

The previously defined scales are ion related and belong to the kinetic scales of the plasma.
Global features of the flow, however, have their own characteristic scales such as the energy containing scale $L_0=2\pi\int (E(k)/k)dk / \int E(k)dk$ where $E(k)$ is the isotropic energy spectrum.
Both fields $\vb{u}_i$ and $\vb{b}$ define characteristic fluctuation velocities $u_0 = \langle\abs{\vb{u}_i}^2\rangle^{1/2}$ and $v_0 = \langle\abs{\vb{b}}^2/4\pi\rho_0 \rangle^{1/2}$ and their corresponding turn-over times through $L_0$.
For simplicity, from here onward we will work with the magnetic field in alfvénic units (i.e., $\vb{b}$ should be understood as $\vb{b}/\sqrt{4\pi\rho_0}$).

Another important parameter is the plasma-$\beta$, defined as the ratio of thermal to magnetic pressure $\beta = 8\pi p_0/\abs{\vb{B_0}}^2 = 2 (p_0/\rho_0)/v_A^2$.
For fluid models such as CHMHD, it is more natural to relate $p_0/\rho_0$ to the sound speed $c_s$ through $c_s^2 = \gamma p_0/\rho_0$.
This choice yields $\beta = (2/\gamma)(c_s/v_A)^2$, which can be understood in terms of the Mach number of the propagating Aflvén waves.
In general, each species has its own plasma-$\beta$ and both contribute to the total pressure such that $\beta = \beta_i+\beta_e$.
For kinetic descriptions it is more common to resort to thermal velocities $v_{\mathrm{th}} = \sqrt{T/m}$, such that $p_{i,0}/\rho_0 = T_i/m_i = v_{\mathrm{th,i}}^2$. 
Therefore, in the hybrid model it is more natural to write $\beta = (v_{\mathrm{th},i}/v_A)^2 + \beta_e$ where $v_{\mathrm{th},i}$ and $\beta_e$ are the relevant parameters for the simulation to set the initial particle distribution and $p_{e,0}$ (see below).

For the CHMHD model, the addition of $\mu$ and $\eta$ give rise to a Reynolds number $R_e = \rho_0u_0L_0/\mu$ and its magnetic counterpart $R_m = v_0L_0/\eta$.
There is also a magnetic Prandlt number $\mathrm{Pr} = \mu/\rho_0\eta$, which we will take equal to $1$ in all our simulations.

\section{Numerical set up}\label{sec:num_set}
\subsection{Hybrid simulations}\label{ssec:hpicsim}

To simulate the hybrid model of Sec.~\ref{ssec:HPICeq} we will use a particle-in-cell method (PIC), which relies on proposing a solution of the form
\begin{equation}
    f_i(\vb{x}, \vb{v}, t) = \sum_{j=1}^{N_p} w_jS\left(\vb{x} - \vb{x}_j(t)\right)\delta \left(\vb{v} - \vb{v}_j(t)\right), \label{eq:exp_fi_pic}
\end{equation}
where $N_p$ is the number of macro-particles, each at $\left(\vb{x}_j(t), \vb{v}_j(t)\right)$ in phase space.
We regard these macro-particles (from now on, simply particles) as a collection of ions bundled together in phase space, occupying a volume determined by the shape function $S$ with well defined velocity.
The amount of ions represented by each particle is determined by its weight $w_j$.
Introducing this in Eq.~\eqref{eq:vlasov_ions} yields Newton's equations of motion for each particle
\begin{equation}
    \dot{\vb{x}}_j = \vb{v}_j, \qquad \dot{\vb{v}}_j = {q_i \over m_i}\left( \vb{E}_j + {\vb{x}_j \over c}\cross\vb{B}_j \right), \label{eq:newton_part}
\end{equation}
where $\vb{E}_j$ and $\vb{B}_j$ are the values of the fields convoluted with the shape function
\begin{equation}
    \vb{E}_j = \int \vb{E}(\vb{x})S\left(\vb{x} - \vb{x}_j\right) \dd\vb{x}, \label{eq:interp_pic}
\end{equation}
and similarly for $\vb{B}_j$.
The shape function is usually obtained by successive convolutions of an initial Dirac's $\delta$ distribution with the so-called 1D top-hat function
\begin{equation}
    \Pi(x) = {1\over\Delta}\left\{
    \begin{matrix}
        1 & \mathrm{if}~|x|<\Delta/2 \\
        0 & \mathrm{otherwise}
    \end{matrix}
    \right.,
\end{equation}
such that $\Pi(\vb{x}) = \Pi(x)\Pi(y)\Pi(z)$ where $\Delta$ is a characteristic macroparticle size, usually chosen to be the grid spacing of the simulation.
The amount of times this convolution is performed determines the order of the shape function $S$, with more convolutions yielding larger particles and therefore smoother density profiles.

Similarly, moments of the ion particle distribution function (such as $n_i$ and $n_i\vb{u}_i$) are obtained by depositing particle properties on the grid.
Applying Eq.~\eqref{eq:exp_fi_pic} into Eq.~\eqref{eq:def_ni}
\begin{equation}
    n_i(\vb{x},t) = \sum_{j=1}^{N_p}  w_jS\left( \vb{x} - \vb{x}_j(t) \right). \label{eq:calc_ni_pic}
\end{equation}

We define the $n$-th order moment of the $\ell$ velocity component of the distribution function as
\begin{equation}
    M_{\ell, n}(\vb{x}, t) = {1\over n_i(\vb{x}, t)}\sum_{j=1}^{N_p} \left[v_{j, \ell}(t)\right]^n w_jS\left( \vb{x} - \vb{x}_j(t) \right), \label{eq:moment_pic}
\end{equation}
where $v_{j, \ell}$ is the $\ell$ component of the velocity $\vb{v}_j$ and thus $u_{i,\ell} = M_{\ell, 1}$.
Centered moments of the distribution $\delta M_{\ell, n}$ can be obtained by replacing $v_\ell$ with $v_\ell-u_{i,\ell}$ in Eq.~\eqref{eq:moment_pic}.
This allows us to anisotropically define a temperature for each direction $T_\ell = \delta M_{\ell, 2}/n_i$ (we drop the mass $m_i$ for simplicity) and higher order statistics such as the kurtosis
\begin{equation}{\label{eq:def_kurt_field}}
    \kappa_{\ell}(\vb{x}, t) = {\delta M_{\ell, 4}(\vb{x}, t) \over T_\ell^2(\vb{x}, t)}.
\end{equation}

Integration is performed using the GHOST code\cite{Mininni2011} with an adapted particle module to treat the macroparticles.
Evolution of Eqs.~\eqref{eq:newton_part} is performed by a Boris scheme modified to fit within the second order Runge-Kutta method used to evolve fields by GHOST (see below).
However, to correctly solve whistler wave dynamic, we add sub-cycling for the magnetic field, performing 10 magnetic field steps per particle step.
We will identify this model as the Hybrid Particle-In-Cell (HPIC).

\subsection{CHMHD simulation}\label{ssec:chmhdsim}

For the compressible Hall magnetohydrodynamic (CHMHD) model, we also use the GHOST code\cite{Mininni2005,Andres2018}.
This code solves the partial diffential equations (PDEs) using a Fourier pseudospectral method to compute spatial derivatives of Eqs. \eqref{eq:continuity}, \eqref{eq:NS} and \eqref{eq:faraday_law_chmhd} in a periodic box.
This scheme ensures exact energy conservation for the continuous-time spatially discrete equations.
Time integration is performed using a second-order Runge-Kutta method and aliasing is removed using the two-thirds truncation method\cite{Orszag1971}, such that the maximum wavenumber resolved is $\kappa\equiv N/3$.
In order to solve the smallest scales, in every simulation we choose $\mu$ and $\eta$ in \eqref{eq:NS} and \eqref{eq:faraday_law_chmhd} such that $\kappa \gtrsim k_d$, where $k_d=(\epsilon_d/\eta^3)^{1/4}$ is the Kolmogorov dissipation wavenumber and $\epsilon_d$ is the energy dissipation rate.

First, we reach a steady turbulent state by using an electromotive forcing $\curl\vb{m}$ (to ensure $\grad\vdot\vb{b} = 0$), which is added to the right-hand side of Eq. \eqref{eq:faraday_law_chmhd}.
This forcing is generated with random phases and constant amplitude in the Fourier $k$-shells $1\leq\abs{\vb{k}}\leq 2$ with a correlation time $\tau_f \approx 4.5\Omega_i^{-1}$.
We perform this process for two cases.
First, a $2.5$D simulation with resolution $\mathrm{N}_x\times \mathrm{N}_y\times \mathrm{N}_z = 512\times512\times1$ where vector fields have all three components but depend only on $x,y$ (i.e., the plane perpendicular to $\vb{B_0}$).
In this simulation, we set $L_{\mathrm{box}} \approx 140 d_i$ such that $k_\mathrm{min}d_i \approx 4.5\times 10^{-2}$ and $k_\mathrm{max}d_i \approx 7.7$, which yields about a whole decade of scales both above and below the ion inertial length $d_i$.
Secondly, a $3$D simulation with reduced resolution along the guide field direction $\mathrm{N}_x\times \mathrm{N}_y\times \mathrm{N}_z = 128\times128\times48$ with $L_{\mathrm{box}} \approx 35 d_i$ such that $k_\mathrm{min}d_i \approx 1.8\times 10^{-1}$ and $k_\mathrm{max}d_i \approx 7.6$.
While this range is considerably smaller, we are mainly interested in checking how the $2.5$D results hold in $3$D.
This and other relevant parameters of both simulations are summarized in Table \ref{tab:stat_vals}.
As described before, both simulations share their forcing correlation time $\tau_f \approx 4.5\Omega_i^{-1}$, magnetic Prandtl number $\mathrm{Pr}=1$ and also $\beta_e = \beta_i \approx 0.47$.


\begin{table}[ht]
\centering
\caption{Global quantities for initial conditions}
\label{tab:stat_vals}
\begin{tabular}{cccccc}
\hline
Run & $\mathrm{N}_x\times \mathrm{N}_y\times \mathrm{N}_z$ & $L_{\text{box}}/d_i$ & $L_0/d_i$ & $v_A/v_0$ & $v_A/u_0$ \\ 
\hline
$2.5$D & $512\times512\times1\hphantom{0}$ & 140 & 84.4 & 3.24 & 4.55 \\ 
$3$D & $128\times128\times48$ & 34.9 & 21.4 & 4.80 & 3.69 \\ 
\hline
\end{tabular}
\caption{Resolution and characteristic scales of the $2.5$D and $3$D stationary states reached with the CHMHD model and used as initial conditions.}
\end{table}

\subsection{Simulations setup}\label{ssec:sim_setup}

Once the stationary state of the CHMHD simulation is reached, particles are introduced in the system.
Particles are uniformly distributed within the domain, forming its own subgrid within each cell.
For the $2.5$D simulation we use $625$ particles per cell (ppc) while for the $3$D simulation we use $512$ ppc.
Given that each cell has the same amount of particles, we emulate the initial density $\rho$ by adjusting the weight of each particle $w_j = \rho(\vb{x}_j, t)/m_i$.
While this process is not perfect, the resulting density $m_in_i$ displays small deviations from $\rho$ (i.e., $\langle (m_in_i - \rho)^2\rangle^{1/2} \leq 1.6\times10^{-2}\rho_0$ in both simulations).
Similarly, the velocity $\vb{v}_j$ of each particle is initialized by sampling a normal distribution with mean $\vb{u}(\vb{x}_j)$ and variance $T_i(\vb{x}_j)/m_i$, with $T_i = m_ip_i/\rho$ and $p_i = p/2$ together with the adiabatic hypothesis of Eq. \eqref{eq:adiabatic}.
Finally, the order of the shape function $S$ is set to 3, yielding smooth density profiles suitable for this low compressibility simulations.

Afterwards, the simulation is split en two cases.
In the first, the CHMHD model continues its evolution and particles evolve following Eq.~\eqref{eq:newton_part}, reacting to the CHMHD fields but unable to affect them (i.e., test particle approximation).
In the second, particles are used as the ions in the HPIC model and evolve as described in Sec. \ref{ssec:hpicsim}, which is self-consistent.
While dissipation is not present, we keep the forcing to ensure both simulations have similar large scale dynamics, even if that injects energy into the HPIC simulations (see below).
All other parameters are kept equal in both cases.

\section{Results} \label{sec:res}

\subsection{2.5D simulations}\label{ssec:res_2D}

\begin{figure}
    \centering
    \includegraphics[width=\linewidth]{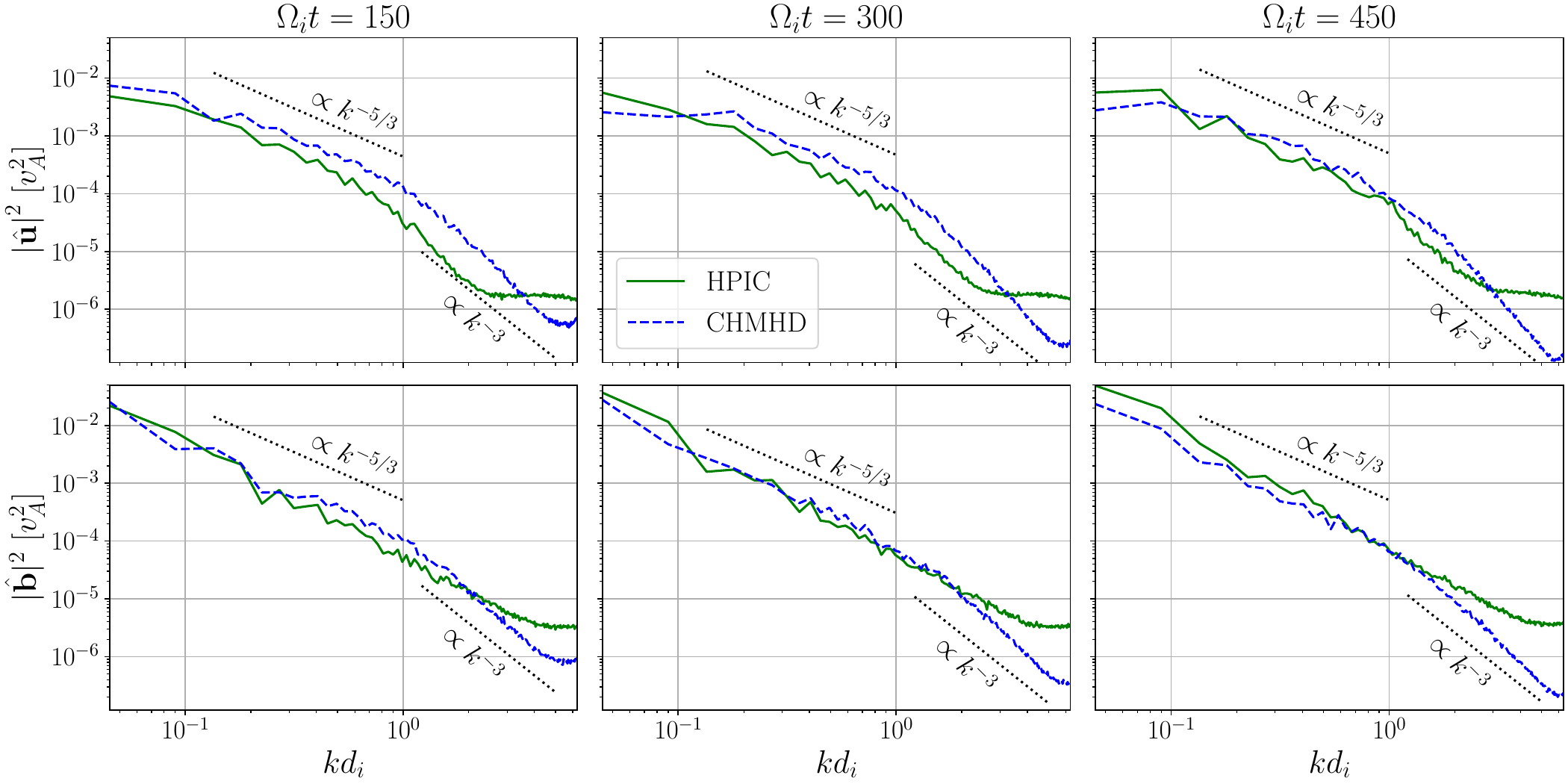}
    \caption{Bulk kinetic (upper) and magnetic (lower) energy spectra at selected times throughout both simulations. A clear Kolmogorov spectrum can be seen for $kd_i\lesssim 1$ in all cases, showing both simulations to have similar large scale flows.} 
    \label{fig:comp_spectra}
\end{figure}

We begin by plotting the bulk kinetic energy and magnetic energy spectra for both simulations at a few chosen times in Fig. \ref{fig:comp_spectra}.
A clear Kolmogorov spectrum can be seen for $kd_i\lesssim 1$ in all cases, giving way to a steeper spectrum, different in each simulation.
In particular, HPIC simulation displays a flat spectrum due to particle shot noise at high $k$, which can be mitigated by increasing the ppc.
This implies that the large scale flow in both simulations is statistically similar, specially in the case of the magnetic field.
Therefore, we can compare these simulations knowing that any difference should be due to kinetic effects and the test particle approximation.

\begin{figure}
    \centering
    \includegraphics[width=\linewidth]{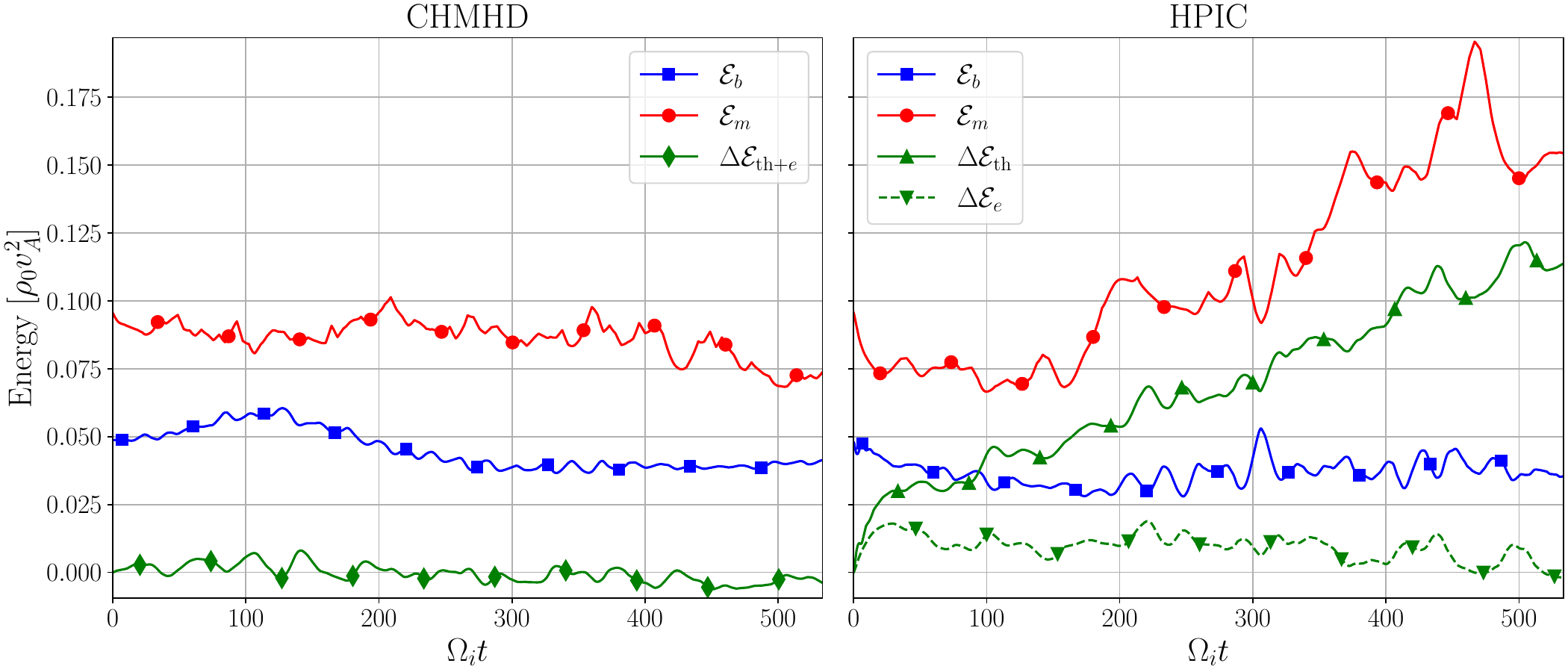}
    \caption{Evolution of the different energies for the CHMHD (left), and HPIC (right) simulations.
    $\mathcal{E}_b$ refers to the bulk kinetic energy, $\Delta \mathcal{E}_{\mathrm{th+e}}$
    is the variation of thermal (ion and electron) energy in the CHMHD case, $\Delta \mathcal{E}_{\mathrm{th}}$ is the variation of thermal ion energy in the HPIC case,  $\Delta \mathcal{E}_e$ is
    the variation of thermal electron energy in the HPIC case and $\mathcal{E}_m$ is the magnetic energy (fluctuating part).   
    While for CHMHD injection is balanced with dissipation and energies remain stationary, for HPIC injected energy is mostly converted into thermal and magnetic energy while the rest remain stationary.} 
    \label{fig:bulk_energies_vs_time}
\end{figure}

Now, we turn to the mean energies in both simulations, whose evolution in time is shown in Fig. \ref{fig:bulk_energies_vs_time}.
We will drop the $\langle\rangle$ from the notation and multiply all energies by $2$ for simplicity.
In order to show all energies in the same scale, we subtract the initial value from the thermal $\mathcal{E}_\mathrm{th}$ and electronic $\mathcal{E}_e$ energies, which start at the same value $\approx 1.42\rho_0v_A^2$.
First, from the left panel we can confirm the reached state to be stationary under the CHMHD model, as all energies mostly fluctuate around their initial values. 
This is not so for the HPIC case, for which there is a fast initial redistribution of energy ($\Omega_i t\lesssim 30$) followed by an interval of $\sim 100\Omega_i^{-1}$ where magnetic and electronic energy are constant but bulk kinetic energy decreases.
After that, magnetic energy starts to increase while bulk kinetic and electronic remain constant.
During all the simulation, thermal energy increases steadily, which hints at some sort of colissionless dissipation.

\begin{figure}
    \centering
    \includegraphics[width=0.5\linewidth]{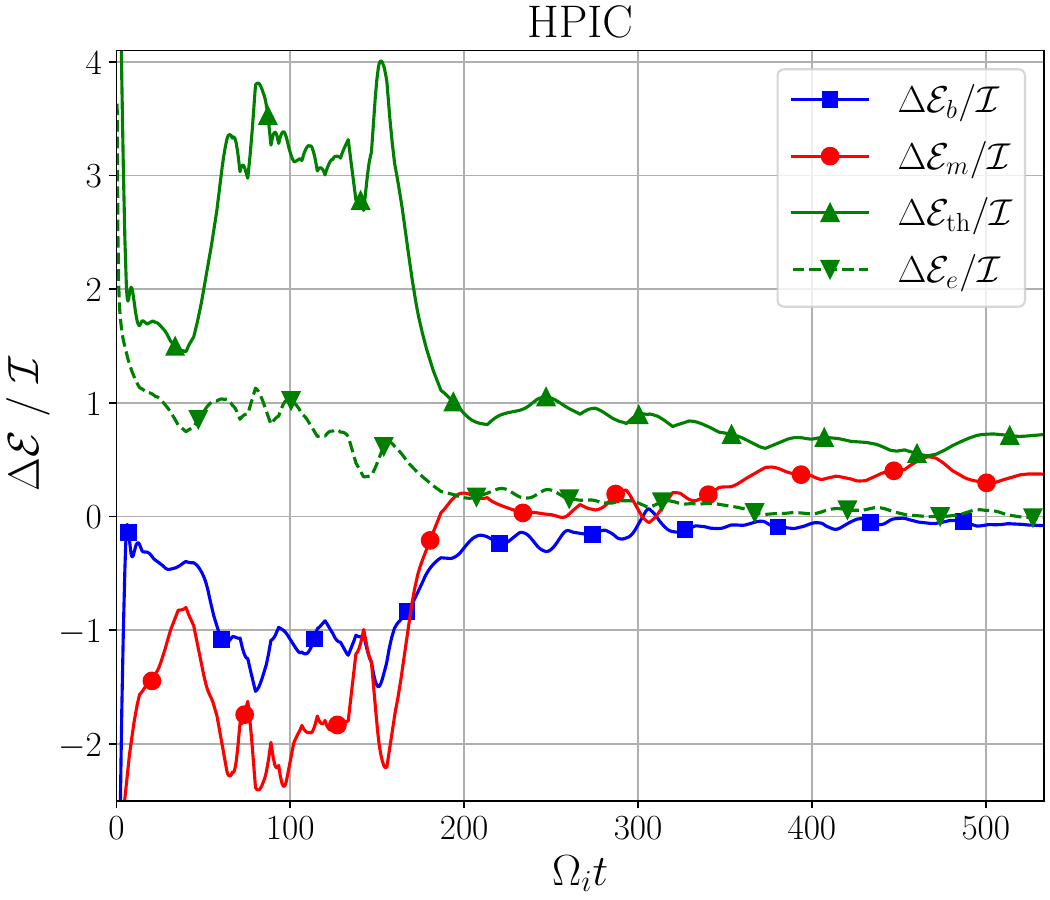}
    \caption{Energy increase normalized by injection for the HPIC case. There is an initial fast energy redistribution for $\Omega_i t\lesssim 30$ followed by an interval where kinetic and magnetic energy are converted into thermal for $30 \lesssim \Omega_i t\lesssim 100$. For $\Omega_i t \gtrsim 100$, injected energy is divided into $\sim 70\%$ thermal and $\sim 30\%$ magnetic energy.}
    \label{fig:inj_vs_time}
\end{figure}

The fact that both magnetic and thermal energies increase until the end of the simulation is due to the constant energy injection by the forcing.
In order to fully understand this process, we calculate the injected energy 
\begin{equation}
    \mathcal{I}(t) = \int_0^t \langle\vb{b}\vdot\curl{\vb{m}}\rangle \dd t',
\end{equation}
where $\langle\vb{b}\vdot\curl{\vb{m}}\rangle$ is the mean energy injection rate.
Figure \ref{fig:inj_vs_time} shows the total variation of each energy $\Delta \mathcal{E}(t) = \mathcal{E}(t) - \mathcal{E}(0)$ normalized by the total energy injected until that moment $\mathcal{I}(t)$.
The three regimes previously discussed are clearly seen here, with some intermediate stages.
As previously discussed, the first regime is simply a fast redistribution of energy, most likely due to a quick reorganization of the system, for which the initial conditions are not fully stable.
During the second stage, $\mathcal{E}_\mathrm{th}$ receives around $3$ times the injected energy, which is only at the expense of $\mathcal{E}_b$ and $\mathcal{E}_m$.
In the last phase, $\mathcal{E}_b$ and $\mathcal{E}_e$ remain constant, such that all injected energy is divided between $\mathcal{E}_m$ ($\sim 30\%$) and $\mathcal{E}_\mathrm{th}$ ($\sim 70\%$).
This points to a mechanism of collisionless dissipation, in which most of the injected energy is converted into thermal energy (heat).

\begin{figure}
    \centering
    \includegraphics[width=\linewidth]{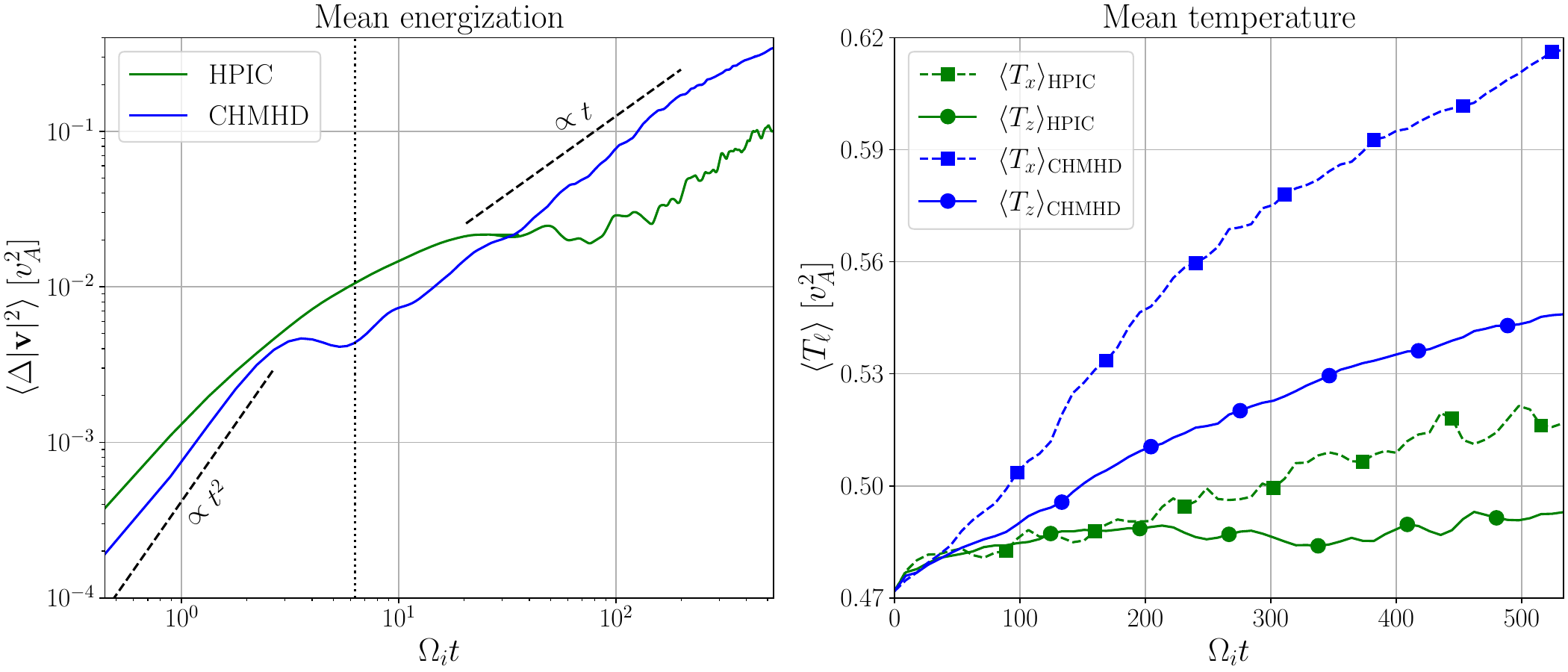}
    \caption{(Left) Mean kinetic energy variation of the particles in both cases as a function of time, showing an initial ballistic increase followed by a a diffusive regime.
    (Right) Mean temperature in each direction as a function of time for both cases, showing about twice the increase of perpendicular component relative to the parallel one.
    Both figures show a higher energy increase for the CHMHD case.}
    \label{fig:energ_vs_time}
\end{figure}

Now we turn to the question of how this heat is distributed and the differences between the test particle CHMHD case and the HPIC case.
In the left panel of Figure \ref{fig:energ_vs_time} we show the total particle energization (equivalent to $\Delta(\mathcal{E}_b+\mathcal{E}_\mathrm{th})$ in the HPIC case) as a function of time.
While initially both display a very similar ballistic range (in velocity space), after a few gyrations test particles (CHMHD case) surpass their self-consistent counterpart (HPIC case).
For $\Omega_i t\gtrsim 100$, consistent with the regime where $\sim70\%$ of injected energy becomes thermal, particles in the HPIC case manage to achieve a slope comparable to the test particles of CHMHD, a diffusive regime.
This energization, however, does not distinguish bulk flow from thermal fluctuations, so in the right panel of Figure \ref{fig:energ_vs_time} we show just the mean thermal fluctuations $T_\ell$ where $\ell$ is the direction, averaged over the whole domain.
As expected, both cases show more perpendicular ($x$ direction, $y$ yields similar results) than parallel ($z$ direction) temperature, in particular $\Delta \langle T_x\rangle \approx 2 \Delta \langle T_z\rangle$.
However, test particles gain $\sim 3$ times more temperature (in both directions) than their self-consistent counterparts, which is almost the same as the total energization of the left panel.
This would imply that almost all particle energization becomes thermal, for both cases.
For the HPIC case, this is expected because Figure \ref{fig:bulk_energies_vs_time} shows the bulk kinetic energy to be roughly constant.
For the CHMHD case, this is not trivial because the test particle bulk velocity $\vb{u}_i$ is not necessarily similar to the fluid velocity $\vb{u}$.
Although not shown here, both fields are very similar at the start of the CHMHD simulation and eventually grow rather different while keeping similar statistical properties such as their mean energy and large scale spectra.

\begin{figure}
    \centering
    \includegraphics[width=\linewidth]{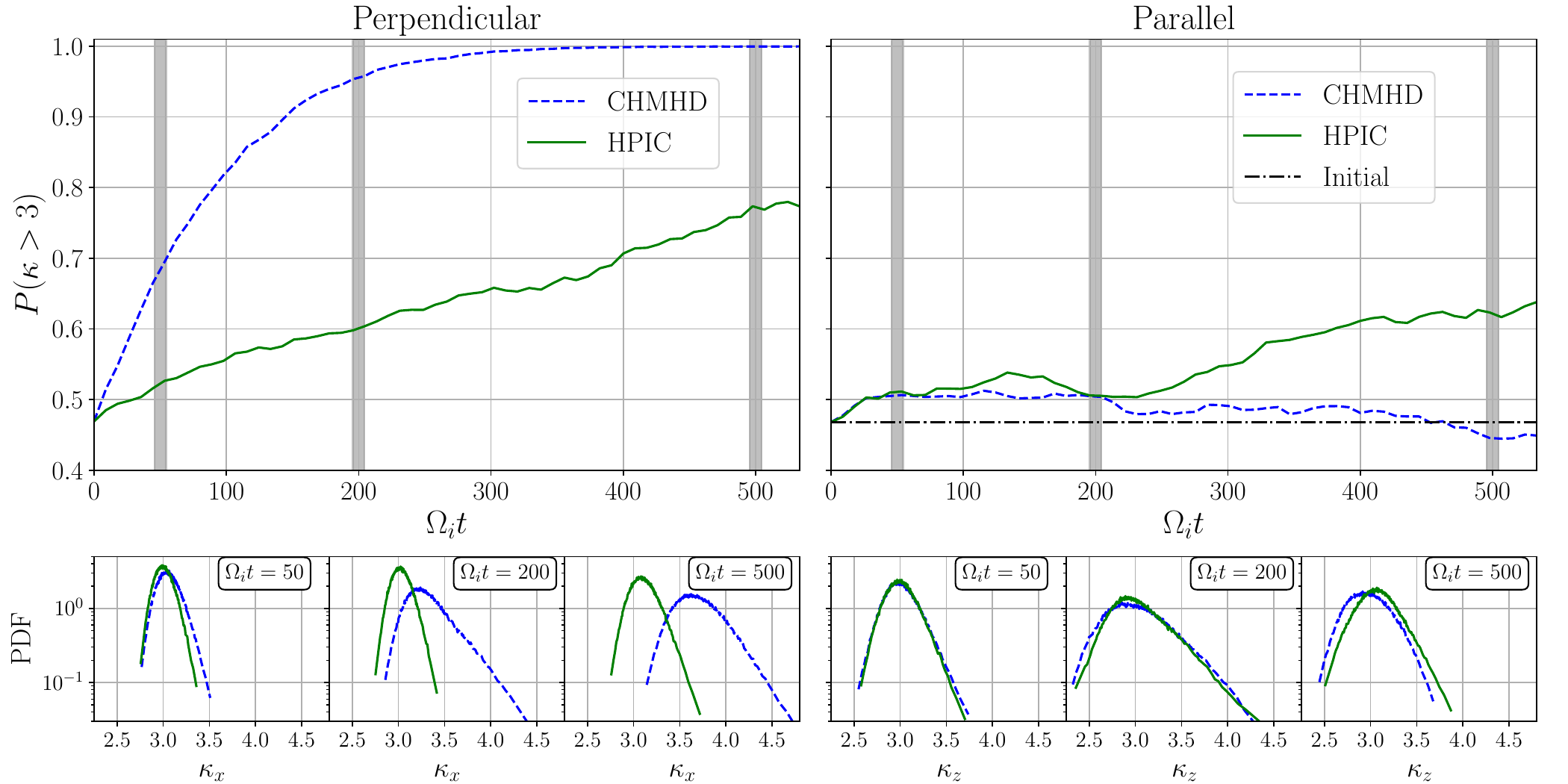}
    \caption{(Top panels) Fraction of grid points with kurtosis $\kappa > 3$ of the velocity PDF for perpendicular (left) and parallel (right) component.
    (Bottom panels) Kurtosis PDFs at different times.}
    \label{fig:kurt_dist_vs_time}
\end{figure}

In order to quantify the production of suprathermal particles, we analyze their velocity distribution.
In particular, we are interested in quantifying how many particles have kinetic energy exceeding what would be expected from a Maxwell-Boltzmann (gaussian) distribution.
Since particles in different regions of the domain have different bulk velocities and temperatures, instead of directly computing the global probability density function (PDF) we propose to calculate the kurtosis field $\kappa_\ell$ as defined in Eq. \eqref{eq:def_kurt_field}, knowing that for a Maxwell-Boltzmann distribution it should be $\kappa_\ell\approx 3$.
Higher values of the kurtosis imply an increased fraction of suprathermal particles in that region of the domain.
In order to analyze the evolution of this field, in Figure \ref{fig:kurt_dist_vs_time} we show the fraction of grid points with $\kappa_\ell > 3$ for each direction along with kurtosis PDFs at some chosen times.
All cases start with $P(\kappa_\ell > 3)\approx 0.47$ because of the inevitable finite size noise of the initial gaussian distribution.
In the perpendicular case, both cases increase in time, with the test particles of the CHMHD case reaching a probability of $1$ rather quickly.
This implies that there is not a single region in the domain without a considerable fraction of suprathermal particles, as can be seen from the PDFs in the lower panels.
The increase in the HPIC case is more modest, but does not seem to stop during the simulation, probably because of the constant injection of energy.
In the parallel direction we observe the opposite, with CHMHD fluctuating around the initial value and, therefore, implying no suprathermal particles.
For HPIC, there seems to be a slight increase, albeit smaller than that of the perpendicular direction.
We see then that the test particle approximation not only overestimates temperature increase, but also generates a much higher (lower) suprathermal population in the perpendicular (parallel) direction.

\begin{figure}
    \centering
    \includegraphics[width=\linewidth]{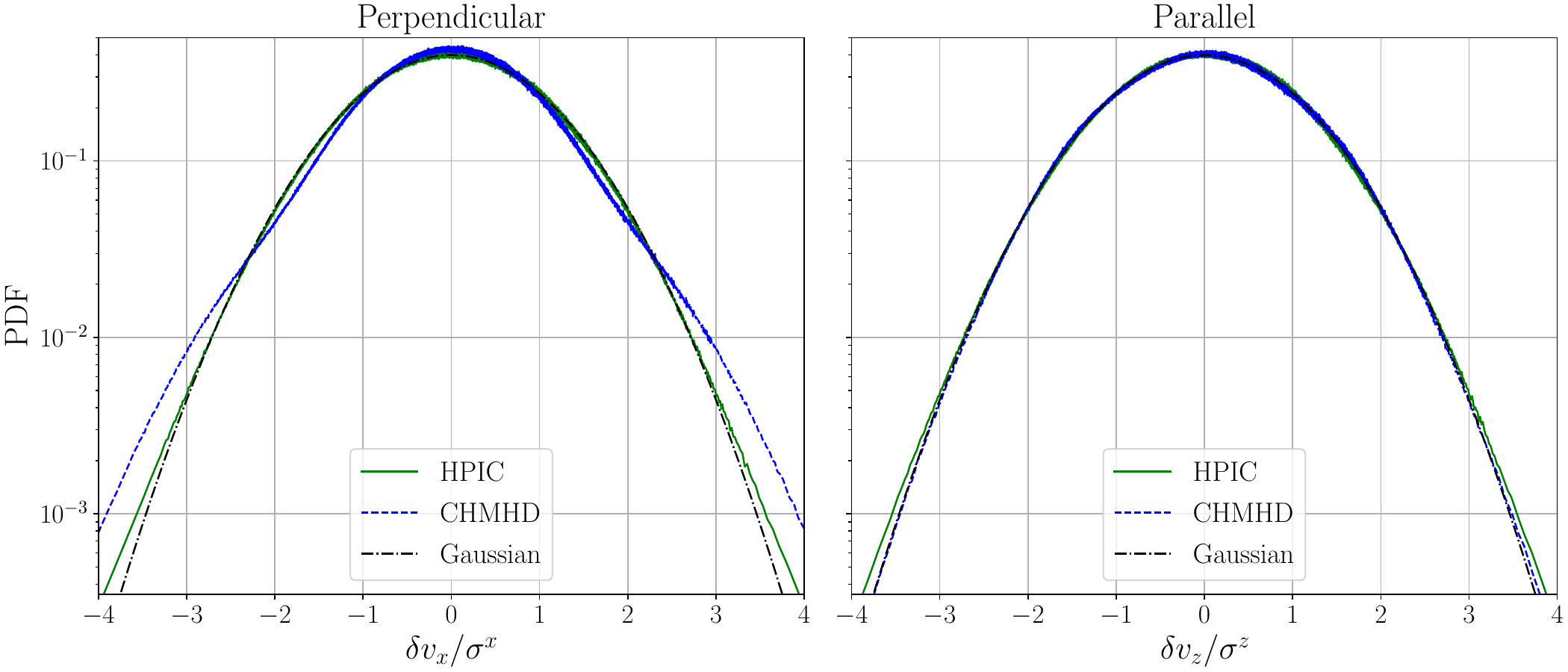}
    \caption{Normalized velocity fluctuation (see text for definition) PDFs at the end of the simulations, for both perpendicular (left), and parallel (right) components.
    The CHMHD case displays heavier tails in the perpendicular direction but it is gaussian in the parallel direction.
    The HPIC case has heavy tails in both directions.} 
    \label{fig:final_deltav_pdf}
\end{figure}

To conclude this analysis, in Figure \ref{fig:final_deltav_pdf} we show the global PDF of the normalized velocities at the end of the simulation.
This normalization is achieved by subtracting the local bulk flow and dividing by the local thermal velocities in each direction such that $\delta \vb{v}_j = \vb{v}_j - \vb{u}(\vb{x}_j)$ and $\sigma^\ell_j = \sqrt{T_\ell(\vb{x}_j)}$.
The previous obervations can be clearly seen here, with both cases displaying heavier tails than the gaussian case in the perpendicular direction, specially so for the CHMHD case.
In the parallel, CHMHD is basically indistinguishable from the gaussian, while HPIC displays slightly heavier tails.
This could simply be a consequence of the $2.5$D approximation, where we disregarded any variation along the parallel direction, but it is however remarkable that the self-consistent particles of HPIC manage to develop some suprathermal properties in spite of that.

\subsection{3D simulations}\label{ssec:res_3D}

\begin{figure}
    \centering
    \includegraphics[width=\linewidth]{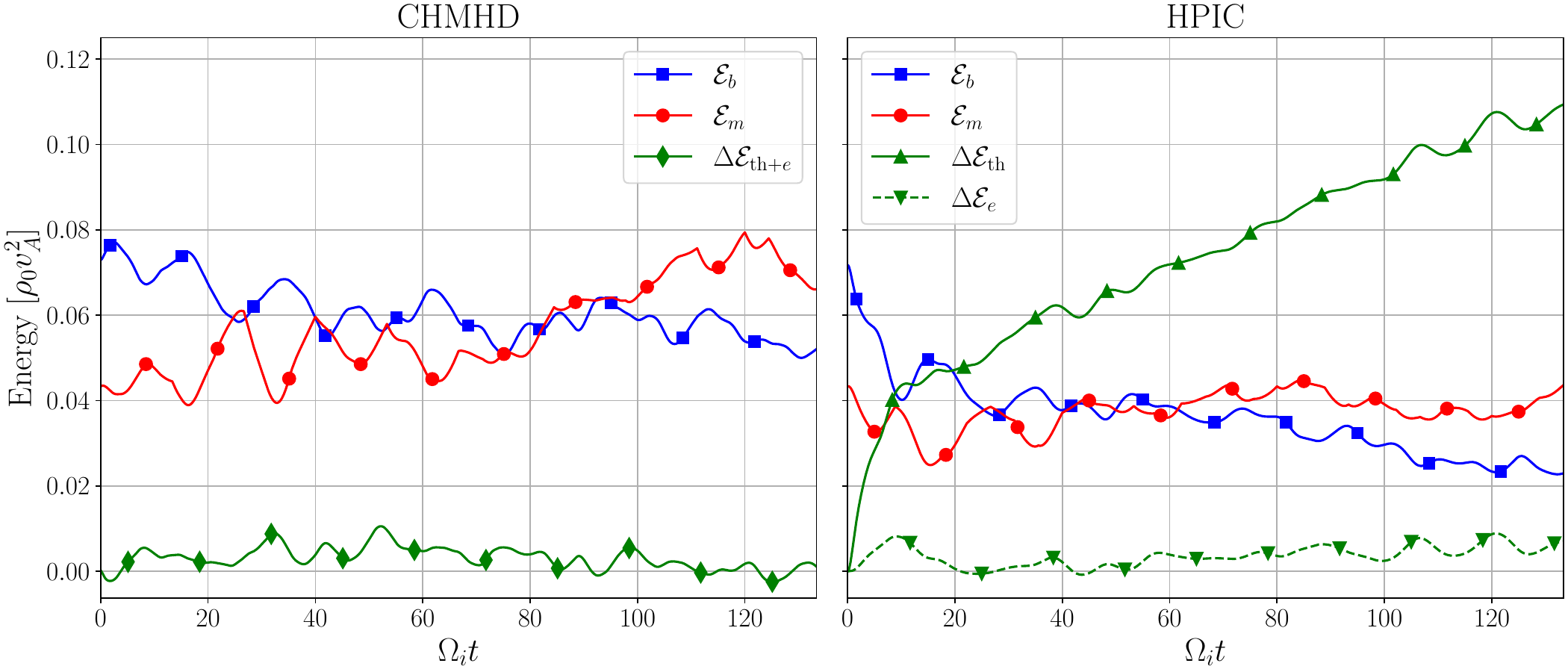}
    \caption{Evolution of the different energies for the CHMHD (left), and HPIC (right) 3D simulations.
    In CHMHD, injection is balanced with dissipation and energies remain stationary, whereas in HPIC, injected energy is mostly converted into thermal energy while the rest remain stationary (in contrast with the 2.5D simulations where also the magnetic energy increases).}
    \label{fig:bulk_energies_vs_time_3d}
\end{figure}

We repeat part of the analysis for a 3D case, with lower resolution due to computational limitations.
The low resolution implies a different separation between scales, in particular between the injection scale and the ion scale and also the duration of the simulation (measured in $\Omega_i^{-1}$) is shorter.
This is bound to change some of the results, so our approach is mostly qualitative.
We begin by showing in Figure \ref{fig:bulk_energies_vs_time_3d} the energy evolution in both cases.
The CHMHD case is quite similar to its $2.5$D counterpart, remaining in a stationary state (although bulk kinetic energy is now higher than magnetic energy).
The HPIC case, however, is quite different, specially at later times.
The initial reorganization can be seen here lasting $\sim 15\Omega_i^{-1}$, but it is then followed by a stationary magnetic energy and a slightly decreasing bulk kinetic energy.
On the other hand, thermal energy increases steadily throughout the whole simulation, which seems to imply that all of the injected energy is transformed into heat.
This could be understood by recalling that in this 3D simulation the injection scale is much closer to the ion scale, thus reducing the MHD cascade range and facilitating its transition to the sub-ion scale, where it is converted into thermal energy. 

\begin{figure}
    \centering
    \includegraphics[width=\linewidth]{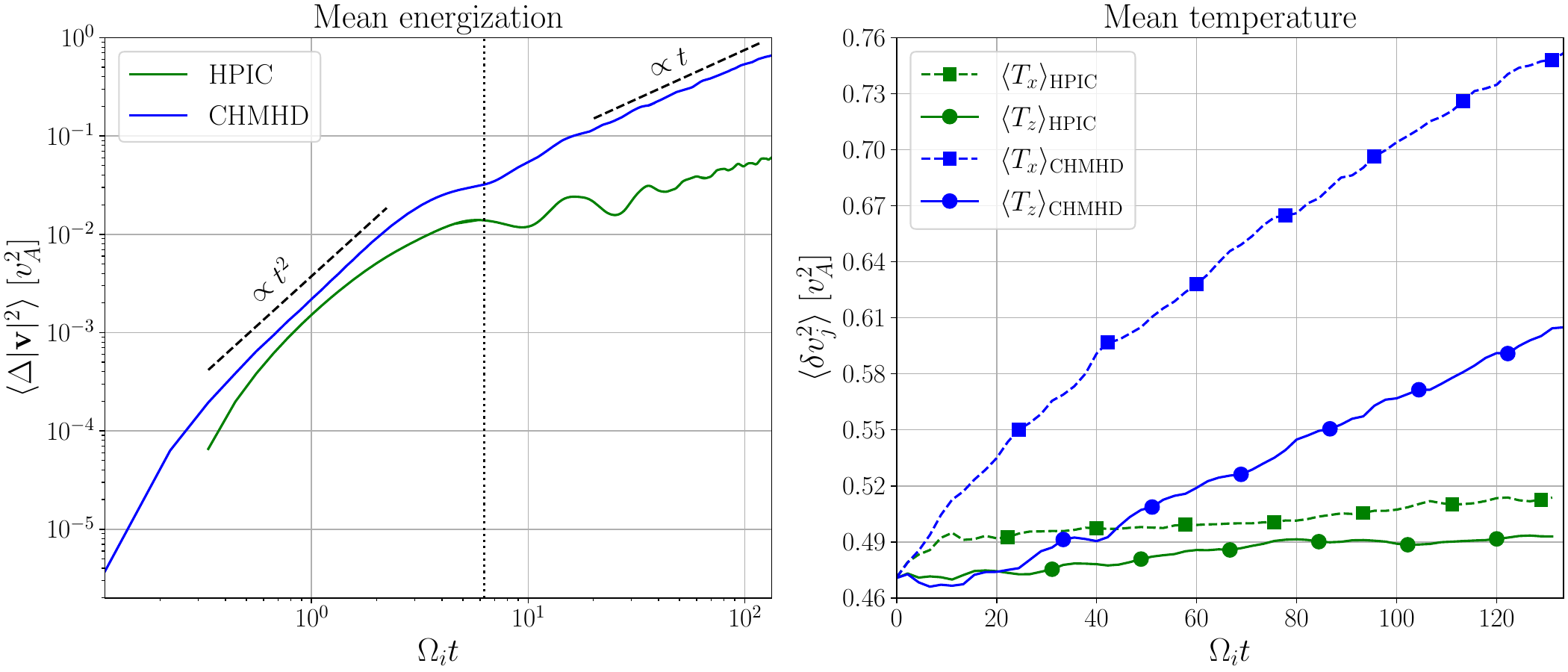}
    \caption{(Left) Mean kinetic energy variation of the particles in both 3D cases (CHMHD and HPIC) as a function of time, showing an  initial ballistic increase followed by a a diffusive regime. (Right) Mean temperature in each direction as a function of time for both cases. Results are similar to Fig. \ref{fig:energ_vs_time}, but here particles of CHMHD gains $\sim 3$ times the amount of energy while for HPIC are very similar.}
    \label{fig:energ_vs_time3d}
\end{figure}

Now, we turn to particle energization, which can be seen in Figure \ref{fig:energ_vs_time3d}.
The initial quadratic increase in left panel is very similar, but the missing points are due to an initial decrease of energy for the self-consistent particles of HPIC.
At later times, HPIC energization seems slower (sub-diffusive) relative to the 2.5D simulation.
More interesting is the mean temperature plot of the right panel, where we can see that temperature increase in HPIC is very similar relative to the 2.5D simulation.
For CHMHD, however, test particle increase their temperatures much more (approximately twice) relative to its 2.5D counterpart and $\sim 6$ times more than the HPIC case.
Relative increases between parallel and perpendicular within each case are mostly the same.
Therefore, we see that the change from 2.5D to 3D has a greater effect on test particles that on self-consistent particles.

\begin{figure}
    \centering
    \includegraphics[width=\linewidth]{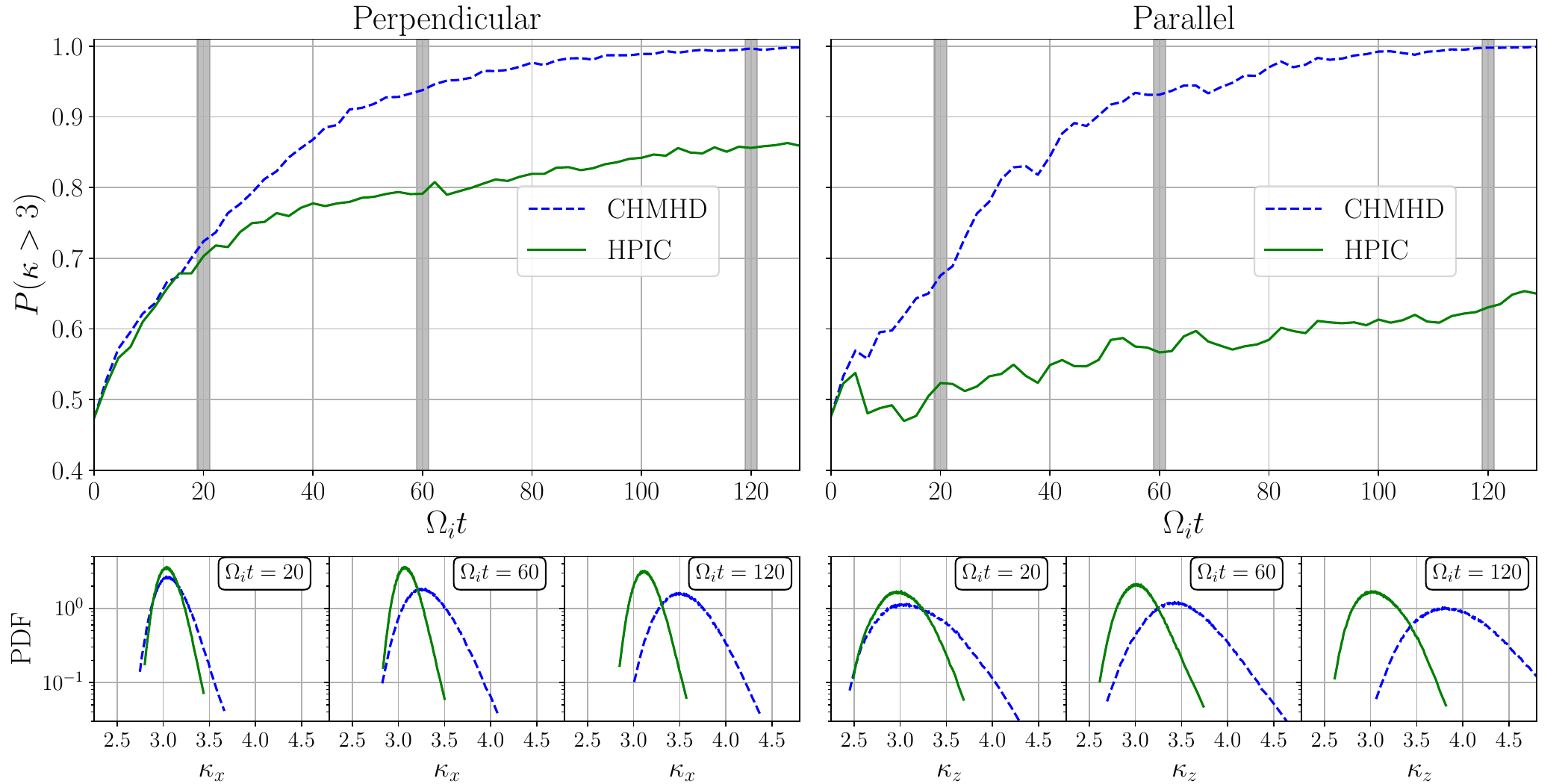}
    \caption{(Top panels) Fraction of grid points where the kurtosis $\kappa > 3$ of the velocity PDF for both parallel (left) and parallel (right) component in the
    3D cases.
    (Bottom panels) Kurtosis PDFs at different times.} 
    \label{fig:kurt_dist_vs_time_3d}
\end{figure}

\begin{figure}
    \centering
    \includegraphics[width=\linewidth]{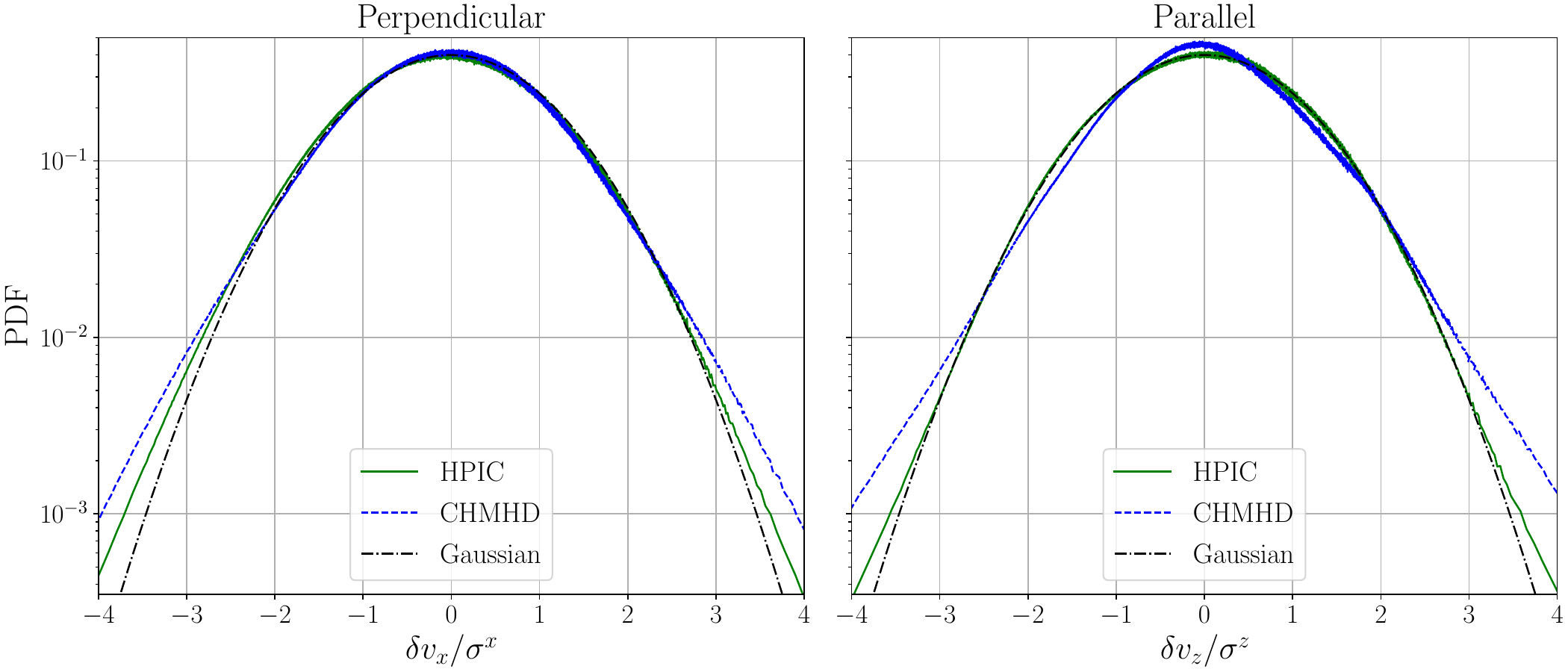}
    \caption{Normalized velocity fluctuation (see text for definition) PDFs at the end of the simulations, for both perpendicular (left), and parallel (right) components
    in the 3D cases.
    The CHMHD case displays heavier tails in both directions relative to the HPIC case, which is also non gaussian.
    For the HPIC case, PDFs are very similar to the 2.5D case of Fig.~\ref{fig:final_deltav_pdf}.}
    \label{fig:final_deltav_pdf_3d}
\end{figure}

We repeat the kurtosis field computation and analysis for the 3D cases. This is shown in Figure \ref{fig:kurt_dist_vs_time_3d}, which displays a number of differences with respect to the 2.5D case in Figure \ref{fig:kurt_dist_vs_time}.
In the perpendicular direction, we see that for 3D both cases are more closely matched and even indistinguishable at earlier times ($\Omega_i t\lesssim 20$).
Near the end of the simulation, CHMHD again reaches a point where the whole domain is populated by suprathermal particles, while HPIC seems to stabilize at $\sim 85\%$, slightly higher than the 2.5D case.
In the parallel direction, CHMHD behaves very similar to the perpendicular direction, in strong contrast to its 2.5D counterpart.
Nonetheless, the HPIC case resemblances its 2.5D version, achieving a comparable level at the end of the simulation.
At a glance, all PDFs in the HPIC case are quite similar for both the 2.5D and 3D cases.
Something similar happens for the normalized velocities of Figure \ref{fig:final_deltav_pdf_3d}, where the HPIC 3D case is almost indistinguishable from the 2.5D case shown in Figure \ref{fig:final_deltav_pdf}, specially so in the parallel component.
On the other hand, the CHMHD case is very different in its parallel component, but less so in the perpendicular. 

\section{Conclusions}\label{sec:disc}

Throughout this work, we have shown the test particle approximation (CHMHD) to overestimate particle heating relative to a more self-consistent formulation (HPIC).
While this should be obvious enough, we dove deep into the issue, giving a quantitative characterization of this overestimation.
Initially, mean temperature is very similar for $\sim 50\Omega_i^{-1}$ for 2.5D (see right panel in Fig. \ref{fig:energ_vs_time}) and $\sim 10\Omega_i^{-1}$ for 3D (see right panel Fig. \ref{fig:energ_vs_time3d}), which on both cases correspond to $\sim 0.13 t_0$ with $t_0 = L_0/u_0$ the large-eddy turnover time.
Afterwards, CHMHD test particles are heated considerable more, specially in the 3D case, while HPIC particles display similar heating rates in 2.5D and 3D.
However, all simulations display $\Delta\langle T_x\rangle \sim 2 \Delta\langle T_z\rangle$ in the end, showing that the test particle approach is able to capture the preferential heating direction.
Coincidentally, comparison between the right panels of Figures \ref{fig:bulk_energies_vs_time} and \ref{fig:bulk_energies_vs_time_3d} shows the importance of the separation of injection and ion scales in collisionless dissipation in self-consistent simulations (HPIC).
As simulations have $k_\mathrm{inj}\approx k_\mathrm{min}$, the 2.5D case has $k_I d_i \approx 4.5\times 10^{-2}$ and $\sim 70\%$ of injected energy dissipated while the 3D case has $k_I d_i \approx 1.8\times 10^{-1}$ and basically all injected energy dissipated.

Suprathermal particle populations differ quite considerably between the CHMHD and HPIC cases.
Figures \ref{fig:kurt_dist_vs_time} and \ref{fig:kurt_dist_vs_time_3d} show that suprathermal particle distributions separate rather quickly, in most cases even faster than mean temperature does.
This implies that, for the test particle case, the velocity distribution develops heavier tails and therefore would have more suprathermal particles relative to the self-consistent one.
In other words, higher order moments are worse captured by the test particle approximation.
Also in this regard, the HPIC simulations are very robust to change from 2.5D to 3D, showing suprathermal particles in both directions but mostly on the perpendicular one.
The CHMHD simulations display this robustness only on the perpendicular direction, while the parallel is mostly gaussian in 2.5D and develops a suprathermal population very similar to the perpendicular direction in 3D.
Therefore, fluctuations along the guide field are necessary for test particles to develop a suprathermal population in that direction but are mostly irrelevant for the self-consistent particles.
This would imply that self-consistent particles manage to exploit parallel energization mechanisms inaccessible to test particles, most likely kinetic in origin.

In astrophysical contexts, we identify particles in the extreme tails of the velocity PDFs as cosmic rays, which due to their high kinetic energy eventually leave the plasma and are replaced by less energetic particles.
As such, an overestimation of the weight of these tails would imply an overestimation of the rate at which cosmic rays are produced.
Our results here show that, even if heating rate is accounted for by normalizing the velocity distributions as in Figures \ref{fig:final_deltav_pdf} and \ref{fig:final_deltav_pdf_3d}, the test particle approximation would predict a higher concentration of cosmic rays.
By removing one of the couplings (i.e., the effect of particles on the fields), we are making it easier for the particles to break free of the plasma fields: they only need enough kinetic energy to make electromagnetic forces negligible.
If the coupling were preserved (as in the self-consistent case) high kinetic energy could not be enough, as a high cosmic ray concentration would in turn generate stronger electromagnetic fields, arresting acceleration.
This arresting seems to manifest not only in lower mean energization but also in a reduced probability of extreme particles (i.e., cosmic rays).
Even with constant energy being injected, the HPIC case seems to develop suprathermal particles only in specific regions of the plasma, while for the CHMHD case they occupy the whole domain rather quickly (see Figures \ref{fig:kurt_dist_vs_time} and \ref{fig:kurt_dist_vs_time_3d}).
Nevertheless the possibility of high energy particle beam formation and the interaction of those beams with the bulk plasma, in the HPIC self-consistent approach could be a path for the detach of a small population of particles akin for the test particle description, which are hard to observe without a considerable increase in the number of particles.
We have presented here the tools and showed a first analysis of results and comparisons and leave for future work those other additional studies.

\begin{acknowledgments}
The authors acknowledge financial support from PIP CONICET grant No.~11220200101752, PICT ANPCyT Grant No.~2018-4298
UBACyT Grant No. 20020220300122BA and from proyecto REMATE of Redes Federales de Alto Impacto, Argentina.
\end{acknowledgments}

\bibliography{main}{}

\providecommand{\noopsort}[1]{}\providecommand{\singleletter}[1]{#1}%
\begin{thebibliography}{44}%
\makeatletter
\providecommand \@ifxundefined [1]{%
 \@ifx{#1\undefined}
}%
\providecommand \@ifnum [1]{%
 \ifnum #1\expandafter \@firstoftwo
 \else \expandafter \@secondoftwo
 \fi
}%
\providecommand \@ifx [1]{%
 \ifx #1\expandafter \@firstoftwo
 \else \expandafter \@secondoftwo
 \fi
}%
\providecommand \natexlab [1]{#1}%
\providecommand \enquote  [1]{``#1''}%
\providecommand \bibnamefont  [1]{#1}%
\providecommand \bibfnamefont [1]{#1}%
\providecommand \citenamefont [1]{#1}%
\providecommand \href@noop [0]{\@secondoftwo}%
\providecommand \href [0]{\begingroup \@sanitize@url \@href}%
\providecommand \@href[1]{\@@startlink{#1}\@@href}%
\providecommand \@@href[1]{\endgroup#1\@@endlink}%
\providecommand \@sanitize@url [0]{\catcode `\\12\catcode `\$12\catcode `\&12\catcode `\#12\catcode `\^12\catcode `\_12\catcode `\%12\relax}%
\providecommand \@@startlink[1]{}%
\providecommand \@@endlink[0]{}%
\providecommand \url  [0]{\begingroup\@sanitize@url \@url }%
\providecommand \@url [1]{\endgroup\@href {#1}{\urlprefix }}%
\providecommand \urlprefix  [0]{URL }%
\providecommand \Eprint [0]{\href }%
\providecommand \doibase [0]{http://dx.doi.org/}%
\providecommand \selectlanguage [0]{\@gobble}%
\providecommand \bibinfo  [0]{\@secondoftwo}%
\providecommand \bibfield  [0]{\@secondoftwo}%
\providecommand \translation [1]{[#1]}%
\providecommand \BibitemOpen [0]{}%
\providecommand \bibitemStop [0]{}%
\providecommand \bibitemNoStop [0]{.\EOS\space}%
\providecommand \EOS [0]{\spacefactor3000\relax}%
\providecommand \BibitemShut  [1]{\csname bibitem#1\endcsname}%
\let\auto@bib@innerbib\@empty
\bibitem [{\citenamefont {Lin}\ and\ \citenamefont {Hudson}(1971)}]{Lin1971}%
  \BibitemOpen
  \bibfield  {author} {\bibinfo {author} {\bibfnamefont {R.~P.}\ \bibnamefont {Lin}}\ and\ \bibinfo {author} {\bibfnamefont {H.~S.}\ \bibnamefont {Hudson}},\ }\href {\doibase 10.1007/bf00150045} {\bibfield  {journal} {\bibinfo  {journal} {Solar Physics}\ }\textbf {\bibinfo {volume} {17}},\ \bibinfo {pages} {412–435} (\bibinfo {year} {1971})}\BibitemShut {NoStop}%
\bibitem [{\citenamefont {Reames}, \citenamefont {Meyer},\ and\ \citenamefont {von Rosenvinge}(1994)}]{Reames1994}%
  \BibitemOpen
  \bibfield  {author} {\bibinfo {author} {\bibfnamefont {D.~V.}\ \bibnamefont {Reames}}, \bibinfo {author} {\bibfnamefont {J.~P.}\ \bibnamefont {Meyer}}, \ and\ \bibinfo {author} {\bibfnamefont {T.~T.}\ \bibnamefont {von Rosenvinge}},\ }\href {\doibase 10.1086/191887} {\bibfield  {journal} {\bibinfo  {journal} {The Astrophysical Journal Supplement Series}\ }\textbf {\bibinfo {volume} {90}},\ \bibinfo {pages} {649} (\bibinfo {year} {1994})}\BibitemShut {NoStop}%
\bibitem [{\citenamefont {Miller}\ \emph {et~al.}(1997)\citenamefont {Miller}, \citenamefont {Cargill}, \citenamefont {Emslie}, \citenamefont {Holman}, \citenamefont {Dennis}, \citenamefont {LaRosa}, \citenamefont {Winglee}, \citenamefont {Benka},\ and\ \citenamefont {Tsuneta}}]{Miller1997}%
  \BibitemOpen
  \bibfield  {author} {\bibinfo {author} {\bibfnamefont {J.~A.}\ \bibnamefont {Miller}}, \bibinfo {author} {\bibfnamefont {P.~J.}\ \bibnamefont {Cargill}}, \bibinfo {author} {\bibfnamefont {A.~G.}\ \bibnamefont {Emslie}}, \bibinfo {author} {\bibfnamefont {G.~D.}\ \bibnamefont {Holman}}, \bibinfo {author} {\bibfnamefont {B.~R.}\ \bibnamefont {Dennis}}, \bibinfo {author} {\bibfnamefont {T.~N.}\ \bibnamefont {LaRosa}}, \bibinfo {author} {\bibfnamefont {R.~M.}\ \bibnamefont {Winglee}}, \bibinfo {author} {\bibfnamefont {S.~G.}\ \bibnamefont {Benka}}, \ and\ \bibinfo {author} {\bibfnamefont {S.}~\bibnamefont {Tsuneta}},\ }\href {\doibase 10.1029/97ja00976} {\bibfield  {journal} {\bibinfo  {journal} {Journal of Geophysical Research: Space Physics}\ }\textbf {\bibinfo {volume} {102}},\ \bibinfo {pages} {14631–14659} (\bibinfo {year} {1997})}\BibitemShut {NoStop}%
\bibitem [{\citenamefont {Desai}\ and\ \citenamefont {Giacalone}(2016)}]{Desai2016}%
  \BibitemOpen
  \bibfield  {author} {\bibinfo {author} {\bibfnamefont {M.}~\bibnamefont {Desai}}\ and\ \bibinfo {author} {\bibfnamefont {J.}~\bibnamefont {Giacalone}},\ }\href {\doibase 10.1007/s41116-016-0002-5} {\bibfield  {journal} {\bibinfo  {journal} {Living Reviews in Solar Physics}\ }\textbf {\bibinfo {volume} {13}} (\bibinfo {year} {2016}),\ 10.1007/s41116-016-0002-5}\BibitemShut {NoStop}%
\bibitem [{\citenamefont {Vlahos}\ \emph {et~al.}(2019)\citenamefont {Vlahos}, \citenamefont {Anastasiadis}, \citenamefont {Papaioannou}, \citenamefont {Kouloumvakos},\ and\ \citenamefont {Isliker}}]{Vlahos2019}%
  \BibitemOpen
  \bibfield  {author} {\bibinfo {author} {\bibfnamefont {L.}~\bibnamefont {Vlahos}}, \bibinfo {author} {\bibfnamefont {A.}~\bibnamefont {Anastasiadis}}, \bibinfo {author} {\bibfnamefont {A.}~\bibnamefont {Papaioannou}}, \bibinfo {author} {\bibfnamefont {A.}~\bibnamefont {Kouloumvakos}}, \ and\ \bibinfo {author} {\bibfnamefont {H.}~\bibnamefont {Isliker}},\ }\href {\doibase 10.1098/rsta.2018.0095} {\bibfield  {journal} {\bibinfo  {journal} {Philosophical Transactions of the Royal Society A: Mathematical, Physical and Engineering Sciences}\ }\textbf {\bibinfo {volume} {377}},\ \bibinfo {pages} {20180095} (\bibinfo {year} {2019})}\BibitemShut {NoStop}%
\bibitem [{\citenamefont {Parker}(1958)}]{Parker1958}%
  \BibitemOpen
  \bibfield  {author} {\bibinfo {author} {\bibfnamefont {E.~N.}\ \bibnamefont {Parker}},\ }\href {\doibase 10.1086/146579} {\bibfield  {journal} {\bibinfo  {journal} {The Astrophysical Journal}\ }\textbf {\bibinfo {volume} {128}},\ \bibinfo {pages} {664} (\bibinfo {year} {1958})}\BibitemShut {NoStop}%
\bibitem [{\citenamefont {Matthaeus}\ \emph {et~al.}(1999)\citenamefont {Matthaeus}, \citenamefont {Zank}, \citenamefont {Smith},\ and\ \citenamefont {Oughton}}]{Matthaeus1999}%
  \BibitemOpen
  \bibfield  {author} {\bibinfo {author} {\bibfnamefont {W.~H.}\ \bibnamefont {Matthaeus}}, \bibinfo {author} {\bibfnamefont {G.~P.}\ \bibnamefont {Zank}}, \bibinfo {author} {\bibfnamefont {C.~W.}\ \bibnamefont {Smith}}, \ and\ \bibinfo {author} {\bibfnamefont {S.}~\bibnamefont {Oughton}},\ }\href {\doibase 10.1103/physrevlett.82.3444} {\bibfield  {journal} {\bibinfo  {journal} {Physical Review Letters}\ }\textbf {\bibinfo {volume} {82}},\ \bibinfo {pages} {3444–3447} (\bibinfo {year} {1999})}\BibitemShut {NoStop}%
\bibitem [{\citenamefont {Hellinger}\ \emph {et~al.}(2013)\citenamefont {Hellinger}, \citenamefont {Tr\'avn\'i\v{c}ek}, \citenamefont {\v{S}t\v{e}p\'an}, \citenamefont {Matteini},\ and\ \citenamefont {Velli}}]{Hellinger2013}%
  \BibitemOpen
  \bibfield  {author} {\bibinfo {author} {\bibfnamefont {P.}~\bibnamefont {Hellinger}}, \bibinfo {author} {\bibfnamefont {P.~M.}\ \bibnamefont {Tr\'avn\'i\v{c}ek}}, \bibinfo {author} {\bibnamefont {\v{S}t\v{e}p\'an}}, \bibinfo {author} {\bibfnamefont {L.}~\bibnamefont {Matteini}}, \ and\ \bibinfo {author} {\bibfnamefont {M.}~\bibnamefont {Velli}},\ }\href {\doibase 10.1002/jgra.50107} {\bibfield  {journal} {\bibinfo  {journal} {Journal of Geophysical Research: Space Physics}\ }\textbf {\bibinfo {volume} {118}},\ \bibinfo {pages} {1351–1365} (\bibinfo {year} {2013})}\BibitemShut {NoStop}%
\bibitem [{\citenamefont {Verscharen}, \citenamefont {Klein},\ and\ \citenamefont {Maruca}(2019)}]{Verscharen2019}%
  \BibitemOpen
  \bibfield  {author} {\bibinfo {author} {\bibfnamefont {D.}~\bibnamefont {Verscharen}}, \bibinfo {author} {\bibfnamefont {K.~G.}\ \bibnamefont {Klein}}, \ and\ \bibinfo {author} {\bibfnamefont {B.~A.}\ \bibnamefont {Maruca}},\ }\href {\doibase 10.1007/s41116-019-0021-0} {\bibfield  {journal} {\bibinfo  {journal} {Living Reviews in Solar Physics}\ }\textbf {\bibinfo {volume} {16}} (\bibinfo {year} {2019}),\ 10.1007/s41116-019-0021-0}\BibitemShut {NoStop}%
\bibitem [{\citenamefont {Matthaeus}\ and\ \citenamefont {Velli}(2011)}]{Matthaeus2011}%
  \BibitemOpen
  \bibfield  {author} {\bibinfo {author} {\bibfnamefont {W.~H.}\ \bibnamefont {Matthaeus}}\ and\ \bibinfo {author} {\bibfnamefont {M.}~\bibnamefont {Velli}},\ }\href {\doibase 10.1007/s11214-011-9793-9} {\bibfield  {journal} {\bibinfo  {journal} {Space Science Reviews}\ }\textbf {\bibinfo {volume} {160}},\ \bibinfo {pages} {145–168} (\bibinfo {year} {2011})}\BibitemShut {NoStop}%
\bibitem [{\citenamefont {Bruno}\ and\ \citenamefont {Carbone}(2013)}]{Bruno2013}%
  \BibitemOpen
  \bibfield  {author} {\bibinfo {author} {\bibfnamefont {R.}~\bibnamefont {Bruno}}\ and\ \bibinfo {author} {\bibfnamefont {V.}~\bibnamefont {Carbone}},\ }\href {\doibase 10.12942/lrsp-2013-2} {\bibfield  {journal} {\bibinfo  {journal} {Living Reviews in Solar Physics}\ }\textbf {\bibinfo {volume} {10}} (\bibinfo {year} {2013}),\ 10.12942/lrsp-2013-2}\BibitemShut {NoStop}%
\bibitem [{\citenamefont {Frisch}(1995)}]{Frisch1995}%
  \BibitemOpen
  \bibfield  {author} {\bibinfo {author} {\bibfnamefont {U.}~\bibnamefont {Frisch}},\ }\href {\doibase 10.1017/cbo9781139170666} {\emph {\bibinfo {title} {Turbulence: The Legacy of A.N. Kolmogorov}}}\ (\bibinfo  {publisher} {Cambridge University Press},\ \bibinfo {year} {1995})\BibitemShut {NoStop}%
\bibitem [{\citenamefont {Sorriso-Valvo}\ \emph {et~al.}(2007)\citenamefont {Sorriso-Valvo}, \citenamefont {Marino}, \citenamefont {Carbone}, \citenamefont {Noullez}, \citenamefont {Lepreti}, \citenamefont {Veltri}, \citenamefont {Bruno}, \citenamefont {Bavassano},\ and\ \citenamefont {Pietropaolo}}]{SorrisoValvo2007}%
  \BibitemOpen
  \bibfield  {author} {\bibinfo {author} {\bibfnamefont {L.}~\bibnamefont {Sorriso-Valvo}}, \bibinfo {author} {\bibfnamefont {R.}~\bibnamefont {Marino}}, \bibinfo {author} {\bibfnamefont {V.}~\bibnamefont {Carbone}}, \bibinfo {author} {\bibfnamefont {A.}~\bibnamefont {Noullez}}, \bibinfo {author} {\bibfnamefont {F.}~\bibnamefont {Lepreti}}, \bibinfo {author} {\bibfnamefont {P.}~\bibnamefont {Veltri}}, \bibinfo {author} {\bibfnamefont {R.}~\bibnamefont {Bruno}}, \bibinfo {author} {\bibfnamefont {B.}~\bibnamefont {Bavassano}}, \ and\ \bibinfo {author} {\bibfnamefont {E.}~\bibnamefont {Pietropaolo}},\ }\href {\doibase 10.1103/physrevlett.99.115001} {\bibfield  {journal} {\bibinfo  {journal} {Physical Review Letters}\ }\textbf {\bibinfo {volume} {99}} (\bibinfo {year} {2007}),\ 10.1103/physrevlett.99.115001}\BibitemShut {NoStop}%
\bibitem [{\citenamefont {Hadid}, \citenamefont {Sahraoui},\ and\ \citenamefont {Galtier}(2017)}]{Hadid2017}%
  \BibitemOpen
  \bibfield  {author} {\bibinfo {author} {\bibfnamefont {L.~Z.}\ \bibnamefont {Hadid}}, \bibinfo {author} {\bibfnamefont {F.}~\bibnamefont {Sahraoui}}, \ and\ \bibinfo {author} {\bibfnamefont {S.}~\bibnamefont {Galtier}},\ }\href {\doibase 10.3847/1538-4357/aa603f} {\bibfield  {journal} {\bibinfo  {journal} {The Astrophysical Journal}\ }\textbf {\bibinfo {volume} {838}},\ \bibinfo {pages} {9} (\bibinfo {year} {2017})}\BibitemShut {NoStop}%
\bibitem [{\citenamefont {Andrés}\ \emph {et~al.}(2019)\citenamefont {Andrés}, \citenamefont {Sahraoui}, \citenamefont {Galtier}, \citenamefont {Hadid}, \citenamefont {Ferrand},\ and\ \citenamefont {Huang}}]{Andres2019}%
  \BibitemOpen
  \bibfield  {author} {\bibinfo {author} {\bibfnamefont {N.}~\bibnamefont {Andrés}}, \bibinfo {author} {\bibfnamefont {F.}~\bibnamefont {Sahraoui}}, \bibinfo {author} {\bibfnamefont {S.}~\bibnamefont {Galtier}}, \bibinfo {author} {\bibfnamefont {L.}~\bibnamefont {Hadid}}, \bibinfo {author} {\bibfnamefont {R.}~\bibnamefont {Ferrand}}, \ and\ \bibinfo {author} {\bibfnamefont {S.}~\bibnamefont {Huang}},\ }\href {\doibase 10.1103/physrevlett.123.245101} {\bibfield  {journal} {\bibinfo  {journal} {Physical Review Letters}\ }\textbf {\bibinfo {volume} {123}} (\bibinfo {year} {2019}),\ 10.1103/physrevlett.123.245101}\BibitemShut {NoStop}%
\bibitem [{\citenamefont {Romanelli}, \citenamefont {Andrés},\ and\ \citenamefont {DiBraccio}(2022)}]{Romanelli2022}%
  \BibitemOpen
  \bibfield  {author} {\bibinfo {author} {\bibfnamefont {N.}~\bibnamefont {Romanelli}}, \bibinfo {author} {\bibfnamefont {N.}~\bibnamefont {Andrés}}, \ and\ \bibinfo {author} {\bibfnamefont {G.~A.}\ \bibnamefont {DiBraccio}},\ }\href {\doibase 10.3847/1538-4357/ac5902} {\bibfield  {journal} {\bibinfo  {journal} {The Astrophysical Journal}\ }\textbf {\bibinfo {volume} {929}},\ \bibinfo {pages} {145} (\bibinfo {year} {2022})}\BibitemShut {NoStop}%
\bibitem [{\citenamefont {{Stix}}(1992)}]{Stix1992}%
  \BibitemOpen
  \bibfield  {author} {\bibinfo {author} {\bibfnamefont {T.~H.}\ \bibnamefont {{Stix}}},\ }\href@noop {} {\emph {\bibinfo {title} {{Waves in plasmas}}}}\ (\bibinfo  {publisher} {American Institute of Physics},\ \bibinfo {year} {1992})\BibitemShut {NoStop}%
\bibitem [{\citenamefont {Bieber}\ \emph {et~al.}(2004)\citenamefont {Bieber}, \citenamefont {Matthaeus}, \citenamefont {Shalchi},\ and\ \citenamefont {Qin}}]{Bieber2004}%
  \BibitemOpen
  \bibfield  {author} {\bibinfo {author} {\bibfnamefont {J.~W.}\ \bibnamefont {Bieber}}, \bibinfo {author} {\bibfnamefont {W.~H.}\ \bibnamefont {Matthaeus}}, \bibinfo {author} {\bibfnamefont {A.}~\bibnamefont {Shalchi}}, \ and\ \bibinfo {author} {\bibfnamefont {G.}~\bibnamefont {Qin}},\ }\href {\doibase 10.1029/2004gl020007} {\bibfield  {journal} {\bibinfo  {journal} {Geophysical Research Letters}\ }\textbf {\bibinfo {volume} {31}} (\bibinfo {year} {2004}),\ 10.1029/2004gl020007}\BibitemShut {NoStop}%
\bibitem [{\citenamefont {Ruffolo}\ \emph {et~al.}(2008)\citenamefont {Ruffolo}, \citenamefont {Chuychai}, \citenamefont {Wongpan}, \citenamefont {Minnie}, \citenamefont {Bieber},\ and\ \citenamefont {Matthaeus}}]{Ruffolo2008}%
  \BibitemOpen
  \bibfield  {author} {\bibinfo {author} {\bibfnamefont {D.}~\bibnamefont {Ruffolo}}, \bibinfo {author} {\bibfnamefont {P.}~\bibnamefont {Chuychai}}, \bibinfo {author} {\bibfnamefont {P.}~\bibnamefont {Wongpan}}, \bibinfo {author} {\bibfnamefont {J.}~\bibnamefont {Minnie}}, \bibinfo {author} {\bibfnamefont {J.~W.}\ \bibnamefont {Bieber}}, \ and\ \bibinfo {author} {\bibfnamefont {W.~H.}\ \bibnamefont {Matthaeus}},\ }\href {\doibase 10.1086/591493} {\bibfield  {journal} {\bibinfo  {journal} {The Astrophysical Journal}\ }\textbf {\bibinfo {volume} {686}},\ \bibinfo {pages} {1231–1244} (\bibinfo {year} {2008})}\BibitemShut {NoStop}%
\bibitem [{\citenamefont {Ruffolo}, \citenamefont {Chuychai},\ and\ \citenamefont {Matthaeus}(2006)}]{Ruffolo2006}%
  \BibitemOpen
  \bibfield  {author} {\bibinfo {author} {\bibfnamefont {D.}~\bibnamefont {Ruffolo}}, \bibinfo {author} {\bibfnamefont {P.}~\bibnamefont {Chuychai}}, \ and\ \bibinfo {author} {\bibfnamefont {W.~H.}\ \bibnamefont {Matthaeus}},\ }\href {\doibase 10.1086/503625} {\bibfield  {journal} {\bibinfo  {journal} {The Astrophysical Journal}\ }\textbf {\bibinfo {volume} {644}},\ \bibinfo {pages} {971–980} (\bibinfo {year} {2006})}\BibitemShut {NoStop}%
\bibitem [{\citenamefont {Minnie}\ \emph {et~al.}(2007)\citenamefont {Minnie}, \citenamefont {Bieber}, \citenamefont {Matthaeus},\ and\ \citenamefont {Burger}}]{Minnie2007}%
  \BibitemOpen
  \bibfield  {author} {\bibinfo {author} {\bibfnamefont {J.}~\bibnamefont {Minnie}}, \bibinfo {author} {\bibfnamefont {J.~W.}\ \bibnamefont {Bieber}}, \bibinfo {author} {\bibfnamefont {W.~H.}\ \bibnamefont {Matthaeus}}, \ and\ \bibinfo {author} {\bibfnamefont {R.~A.}\ \bibnamefont {Burger}},\ }\href {\doibase 10.1086/518765} {\bibfield  {journal} {\bibinfo  {journal} {The Astrophysical Journal}\ }\textbf {\bibinfo {volume} {663}},\ \bibinfo {pages} {1049–1054} (\bibinfo {year} {2007})}\BibitemShut {NoStop}%
\bibitem [{\citenamefont {Dalena}\ \emph {et~al.}(2012)\citenamefont {Dalena}, \citenamefont {Greco}, \citenamefont {Rappazzo}, \citenamefont {Mace},\ and\ \citenamefont {Matthaeus}}]{Dalena2012}%
  \BibitemOpen
  \bibfield  {author} {\bibinfo {author} {\bibfnamefont {S.}~\bibnamefont {Dalena}}, \bibinfo {author} {\bibfnamefont {A.}~\bibnamefont {Greco}}, \bibinfo {author} {\bibfnamefont {A.~F.}\ \bibnamefont {Rappazzo}}, \bibinfo {author} {\bibfnamefont {R.~L.}\ \bibnamefont {Mace}}, \ and\ \bibinfo {author} {\bibfnamefont {W.~H.}\ \bibnamefont {Matthaeus}},\ }\href {\doibase 10.1103/PhysRevE.86.016402} {\bibfield  {journal} {\bibinfo  {journal} {Phys. Rev. E}\ }\textbf {\bibinfo {volume} {86}},\ \bibinfo {pages} {016402} (\bibinfo {year} {2012})}\BibitemShut {NoStop}%
\bibitem [{\citenamefont {Tautz}\ and\ \citenamefont {Dosch}(2013)}]{Tautz2013}%
  \BibitemOpen
  \bibfield  {author} {\bibinfo {author} {\bibfnamefont {R.~C.}\ \bibnamefont {Tautz}}\ and\ \bibinfo {author} {\bibfnamefont {A.}~\bibnamefont {Dosch}},\ }\href {\doibase 10.1063/1.4789861} {\bibfield  {journal} {\bibinfo  {journal} {Physics of Plasmas}\ }\textbf {\bibinfo {volume} {20}},\ \bibinfo {pages} {022302} (\bibinfo {year} {2013})},\ \Eprint {http://arxiv.org/abs/https://doi.org/10.1063/1.4789861} {https://doi.org/10.1063/1.4789861} \BibitemShut {NoStop}%
\bibitem [{\citenamefont {{Dalena}}\ \emph {et~al.}(2014)\citenamefont {{Dalena}}, \citenamefont {{Rappazzo}}, \citenamefont {{Dmitruk}}, \citenamefont {{Greco}},\ and\ \citenamefont {{Matthaeus}}}]{Dalena2014}%
  \BibitemOpen
  \bibfield  {author} {\bibinfo {author} {\bibfnamefont {S.}~\bibnamefont {{Dalena}}}, \bibinfo {author} {\bibfnamefont {A.~F.}\ \bibnamefont {{Rappazzo}}}, \bibinfo {author} {\bibfnamefont {P.}~\bibnamefont {{Dmitruk}}}, \bibinfo {author} {\bibfnamefont {A.}~\bibnamefont {{Greco}}}, \ and\ \bibinfo {author} {\bibfnamefont {W.~H.}\ \bibnamefont {{Matthaeus}}},\ }\href {\doibase 10.1088/0004-637X/783/2/143} {\bibfield  {journal} {\bibinfo  {journal} {ApJ}\ }\textbf {\bibinfo {volume} {783}},\ \bibinfo {eid} {143} (\bibinfo {year} {2014})},\ \Eprint {http://arxiv.org/abs/1402.3745} {arXiv:1402.3745 [astro-ph.SR]} \BibitemShut {NoStop}%
\bibitem [{\citenamefont {Dmitruk}, \citenamefont {Matthaeus},\ and\ \citenamefont {Seenu}(2004)}]{Dmitruk2004}%
  \BibitemOpen
  \bibfield  {author} {\bibinfo {author} {\bibfnamefont {P.}~\bibnamefont {Dmitruk}}, \bibinfo {author} {\bibfnamefont {W.~H.}\ \bibnamefont {Matthaeus}}, \ and\ \bibinfo {author} {\bibfnamefont {N.}~\bibnamefont {Seenu}},\ }\href {\doibase 10.1086/425301} {\bibfield  {journal} {\bibinfo  {journal} {The Astrophysical Journal}\ }\textbf {\bibinfo {volume} {617}},\ \bibinfo {pages} {667} (\bibinfo {year} {2004})}\BibitemShut {NoStop}%
\bibitem [{\citenamefont {Dmitruk}\ and\ \citenamefont {Matthaeus}(2006)}]{dmitruk2006b}%
  \BibitemOpen
  \bibfield  {author} {\bibinfo {author} {\bibfnamefont {P.}~\bibnamefont {Dmitruk}}\ and\ \bibinfo {author} {\bibfnamefont {W.~H.}\ \bibnamefont {Matthaeus}},\ }\href@noop {} {\bibfield  {journal} {\bibinfo  {journal} {Physics of Plasmas}\ }\textbf {\bibinfo {volume} {13}},\ \bibinfo {pages} {042307} (\bibinfo {year} {2006})}\BibitemShut {NoStop}%
\bibitem [{\citenamefont {{Gonz{\'a}lez}}\ \emph {et~al.}(2016)\citenamefont {{Gonz{\'a}lez}}, \citenamefont {{Dmitruk}}, \citenamefont {{Mininni}},\ and\ \citenamefont {{Matthaeus}}}]{Gonzalez2016}%
  \BibitemOpen
  \bibfield  {author} {\bibinfo {author} {\bibfnamefont {C.~A.}\ \bibnamefont {{Gonz{\'a}lez}}}, \bibinfo {author} {\bibfnamefont {P.}~\bibnamefont {{Dmitruk}}}, \bibinfo {author} {\bibfnamefont {P.~D.}\ \bibnamefont {{Mininni}}}, \ and\ \bibinfo {author} {\bibfnamefont {W.~H.}\ \bibnamefont {{Matthaeus}}},\ }\href {\doibase 10.1063/1.4960681} {\bibfield  {journal} {\bibinfo  {journal} {Physics of Plasmas}\ }\textbf {\bibinfo {volume} {23}},\ \bibinfo {eid} {082305} (\bibinfo {year} {2016})},\ \Eprint {http://arxiv.org/abs/1605.02811} {arXiv:1605.02811 [physics.plasm-ph]} \BibitemShut {NoStop}%
\bibitem [{\citenamefont {Gonz{\'{a}}lez}\ \emph {et~al.}(2017)\citenamefont {Gonz{\'{a}}lez}, \citenamefont {Dmitruk}, \citenamefont {Mininni},\ and\ \citenamefont {Matthaeus}}]{Gonzalez_2017}%
  \BibitemOpen
  \bibfield  {author} {\bibinfo {author} {\bibfnamefont {C.~A.}\ \bibnamefont {Gonz{\'{a}}lez}}, \bibinfo {author} {\bibfnamefont {P.}~\bibnamefont {Dmitruk}}, \bibinfo {author} {\bibfnamefont {P.~D.}\ \bibnamefont {Mininni}}, \ and\ \bibinfo {author} {\bibfnamefont {W.~H.}\ \bibnamefont {Matthaeus}},\ }\href {\doibase 10.3847/1538-4357/aa8c02} {\bibfield  {journal} {\bibinfo  {journal} {The Astrophysical Journal}\ }\textbf {\bibinfo {volume} {850}},\ \bibinfo {pages} {19} (\bibinfo {year} {2017})}\BibitemShut {NoStop}%
\bibitem [{\citenamefont {Pugliese}\ and\ \citenamefont {Dmitruk}(2022)}]{pugliese2022}%
  \BibitemOpen
  \bibfield  {author} {\bibinfo {author} {\bibfnamefont {F.}~\bibnamefont {Pugliese}}\ and\ \bibinfo {author} {\bibfnamefont {P.}~\bibnamefont {Dmitruk}},\ }\href {\doibase 10.3847/1538-4357/ac5abe} {\bibfield  {journal} {\bibinfo  {journal} {Astrophysical Journal}\ }\textbf {\bibinfo {volume} {929}},\ \bibinfo {pages} {1538} (\bibinfo {year} {2022})}\BibitemShut {NoStop}%
\bibitem [{\citenamefont {Pezzi}, \citenamefont {Blasi},\ and\ \citenamefont {Matthaeus}(2022)}]{Pezzi2022}%
  \BibitemOpen
  \bibfield  {author} {\bibinfo {author} {\bibfnamefont {O.}~\bibnamefont {Pezzi}}, \bibinfo {author} {\bibfnamefont {P.}~\bibnamefont {Blasi}}, \ and\ \bibinfo {author} {\bibfnamefont {W.~H.}\ \bibnamefont {Matthaeus}},\ }\href {\doibase 10.3847/1538-4357/ac5332} {\bibfield  {journal} {\bibinfo  {journal} {The Astrophysical Journal}\ }\textbf {\bibinfo {volume} {928}},\ \bibinfo {pages} {25} (\bibinfo {year} {2022})}\BibitemShut {NoStop}%
\bibitem [{\citenamefont {Balzarini}, \citenamefont {Pugliese},\ and\ \citenamefont {Dmitruk}(2022)}]{Balzarini2022}%
  \BibitemOpen
  \bibfield  {author} {\bibinfo {author} {\bibfnamefont {B.}~\bibnamefont {Balzarini}}, \bibinfo {author} {\bibfnamefont {F.}~\bibnamefont {Pugliese}}, \ and\ \bibinfo {author} {\bibfnamefont {P.}~\bibnamefont {Dmitruk}},\ }\href {\doibase 10.1063/5.0117847} {\bibfield  {journal} {\bibinfo  {journal} {Physics of Plasmas}\ }\textbf {\bibinfo {volume} {29}},\ \bibinfo {pages} {122303} (\bibinfo {year} {2022})},\ \Eprint {http://arxiv.org/abs/https://doi.org/10.1063/5.0117847} {https://doi.org/10.1063/5.0117847} \BibitemShut {NoStop}%
\bibitem [{\citenamefont {Greco}, \citenamefont {Artemyev},\ and\ \citenamefont {Zimbardo}(2014)}]{Greco2014}%
  \BibitemOpen
  \bibfield  {author} {\bibinfo {author} {\bibfnamefont {A.}~\bibnamefont {Greco}}, \bibinfo {author} {\bibfnamefont {A.}~\bibnamefont {Artemyev}}, \ and\ \bibinfo {author} {\bibfnamefont {G.}~\bibnamefont {Zimbardo}},\ }\href {\doibase https://doi.org/10.1002/2014JA020421} {\bibfield  {journal} {\bibinfo  {journal} {Journal of Geophysical Research: Space Physics}\ }\textbf {\bibinfo {volume} {119}},\ \bibinfo {pages} {8929} (\bibinfo {year} {2014})},\ \Eprint {http://arxiv.org/abs/https://agupubs.onlinelibrary.wiley.com/doi/pdf/10.1002/2014JA020421} {https://agupubs.onlinelibrary.wiley.com/doi/pdf/10.1002/2014JA020421} \BibitemShut {NoStop}%
\bibitem [{\citenamefont {Lemoine}(2021)}]{Lemoine2021}%
  \BibitemOpen
  \bibfield  {author} {\bibinfo {author} {\bibfnamefont {M.}~\bibnamefont {Lemoine}},\ }\href {\doibase 10.1103/physrevd.104.063020} {\bibfield  {journal} {\bibinfo  {journal} {Physical Review D}\ }\textbf {\bibinfo {volume} {104}} (\bibinfo {year} {2021}),\ 10.1103/physrevd.104.063020}\BibitemShut {NoStop}%
\bibitem [{\citenamefont {Pugliese}\ \emph {et~al.}(2023)\citenamefont {Pugliese}, \citenamefont {Brodiano}, \citenamefont {Andrés},\ and\ \citenamefont {Dmitruk}}]{pugliese2023}%
  \BibitemOpen
  \bibfield  {author} {\bibinfo {author} {\bibfnamefont {F.}~\bibnamefont {Pugliese}}, \bibinfo {author} {\bibfnamefont {M.}~\bibnamefont {Brodiano}}, \bibinfo {author} {\bibfnamefont {N.}~\bibnamefont {Andrés}}, \ and\ \bibinfo {author} {\bibfnamefont {P.}~\bibnamefont {Dmitruk}},\ }\href {\doibase 10.3847/1538-4357/ad055b} {\bibfield  {journal} {\bibinfo  {journal} {The Astrophysical Journal}\ }\textbf {\bibinfo {volume} {959}},\ \bibinfo {pages} {28} (\bibinfo {year} {2023})}\BibitemShut {NoStop}%
\bibitem [{\citenamefont {Servidio}\ \emph {et~al.}(2017)\citenamefont {Servidio}, \citenamefont {Chasapis}, \citenamefont {Matthaeus}, \citenamefont {Perrone}, \citenamefont {Valentini}, \citenamefont {Parashar}, \citenamefont {Veltri}, \citenamefont {Gershman}, \citenamefont {Russell}, \citenamefont {Giles}, \citenamefont {Fuselier}, \citenamefont {Phan},\ and\ \citenamefont {Burch}}]{Servidio2017}%
  \BibitemOpen
  \bibfield  {author} {\bibinfo {author} {\bibfnamefont {S.}~\bibnamefont {Servidio}}, \bibinfo {author} {\bibfnamefont {A.}~\bibnamefont {Chasapis}}, \bibinfo {author} {\bibfnamefont {W.}~\bibnamefont {Matthaeus}}, \bibinfo {author} {\bibfnamefont {D.}~\bibnamefont {Perrone}}, \bibinfo {author} {\bibfnamefont {F.}~\bibnamefont {Valentini}}, \bibinfo {author} {\bibfnamefont {T.}~\bibnamefont {Parashar}}, \bibinfo {author} {\bibfnamefont {P.}~\bibnamefont {Veltri}}, \bibinfo {author} {\bibfnamefont {D.}~\bibnamefont {Gershman}}, \bibinfo {author} {\bibfnamefont {C.}~\bibnamefont {Russell}}, \bibinfo {author} {\bibfnamefont {B.}~\bibnamefont {Giles}}, \bibinfo {author} {\bibfnamefont {S.}~\bibnamefont {Fuselier}}, \bibinfo {author} {\bibfnamefont {T.}~\bibnamefont {Phan}}, \ and\ \bibinfo {author} {\bibfnamefont {J.}~\bibnamefont {Burch}},\ }\href {\doibase 10.1103/physrevlett.119.205101} {\bibfield  {journal} {\bibinfo  {journal} {Physical Review Letters}\ }\textbf {\bibinfo {volume} {119}} (\bibinfo {year}
  {2017}),\ 10.1103/physrevlett.119.205101}\BibitemShut {NoStop}%
\bibitem [{\citenamefont {Howes}(2017)}]{Howes2017}%
  \BibitemOpen
  \bibfield  {author} {\bibinfo {author} {\bibfnamefont {G.~G.}\ \bibnamefont {Howes}},\ }\href {\doibase 10.1063/1.4983993} {\bibfield  {journal} {\bibinfo  {journal} {Physics of Plasmas}\ }\textbf {\bibinfo {volume} {24}} (\bibinfo {year} {2017}),\ 10.1063/1.4983993}\BibitemShut {NoStop}%
\bibitem [{\citenamefont {Muñoz}\ \emph {et~al.}(2018)\citenamefont {Muñoz}, \citenamefont {Jain}, \citenamefont {Kilian},\ and\ \citenamefont {B\"{u}chner}}]{Muñoz2018}%
  \BibitemOpen
  \bibfield  {author} {\bibinfo {author} {\bibfnamefont {P.}~\bibnamefont {Muñoz}}, \bibinfo {author} {\bibfnamefont {N.}~\bibnamefont {Jain}}, \bibinfo {author} {\bibfnamefont {P.}~\bibnamefont {Kilian}}, \ and\ \bibinfo {author} {\bibfnamefont {J.}~\bibnamefont {B\"{u}chner}},\ }\href {\doibase 10.1016/j.cpc.2017.10.012} {\bibfield  {journal} {\bibinfo  {journal} {Computer Physics Communications}\ }\textbf {\bibinfo {volume} {224}},\ \bibinfo {pages} {245–264} (\bibinfo {year} {2018})}\BibitemShut {NoStop}%
\bibitem [{\citenamefont {Birn}\ \emph {et~al.}(2001)\citenamefont {Birn}, \citenamefont {Drake}, \citenamefont {Shay}, \citenamefont {Rogers}, \citenamefont {Denton}, \citenamefont {Hesse}, \citenamefont {Kuznetsova}, \citenamefont {Ma}, \citenamefont {Bhattacharjee}, \citenamefont {Otto},\ and\ \citenamefont {Pritchett}}]{Birn2001}%
  \BibitemOpen
  \bibfield  {author} {\bibinfo {author} {\bibfnamefont {J.}~\bibnamefont {Birn}}, \bibinfo {author} {\bibfnamefont {J.~F.}\ \bibnamefont {Drake}}, \bibinfo {author} {\bibfnamefont {M.~A.}\ \bibnamefont {Shay}}, \bibinfo {author} {\bibfnamefont {B.~N.}\ \bibnamefont {Rogers}}, \bibinfo {author} {\bibfnamefont {R.~E.}\ \bibnamefont {Denton}}, \bibinfo {author} {\bibfnamefont {M.}~\bibnamefont {Hesse}}, \bibinfo {author} {\bibfnamefont {M.}~\bibnamefont {Kuznetsova}}, \bibinfo {author} {\bibfnamefont {Z.~W.}\ \bibnamefont {Ma}}, \bibinfo {author} {\bibfnamefont {A.}~\bibnamefont {Bhattacharjee}}, \bibinfo {author} {\bibfnamefont {A.}~\bibnamefont {Otto}}, \ and\ \bibinfo {author} {\bibfnamefont {P.~L.}\ \bibnamefont {Pritchett}},\ }\href {\doibase 10.1029/1999ja900449} {\bibfield  {journal} {\bibinfo  {journal} {Journal of Geophysical Research: Space Physics}\ }\textbf {\bibinfo {volume} {106}},\ \bibinfo {pages} {3715–3719} (\bibinfo {year} {2001})}\BibitemShut {NoStop}%
\bibitem [{\citenamefont {Le}\ \emph {et~al.}(2016)\citenamefont {Le}, \citenamefont {Daughton}, \citenamefont {Karimabadi},\ and\ \citenamefont {Egedal}}]{Le2016}%
  \BibitemOpen
  \bibfield  {author} {\bibinfo {author} {\bibfnamefont {A.}~\bibnamefont {Le}}, \bibinfo {author} {\bibfnamefont {W.}~\bibnamefont {Daughton}}, \bibinfo {author} {\bibfnamefont {H.}~\bibnamefont {Karimabadi}}, \ and\ \bibinfo {author} {\bibfnamefont {J.}~\bibnamefont {Egedal}},\ }\href {\doibase 10.1063/1.4943893} {\bibfield  {journal} {\bibinfo  {journal} {Physics of Plasmas}\ }\textbf {\bibinfo {volume} {23}} (\bibinfo {year} {2016}),\ 10.1063/1.4943893}\BibitemShut {NoStop}%
\bibitem [{\citenamefont {González}, \citenamefont {Innocenti},\ and\ \citenamefont {Tenerani}(2023)}]{Gonzlez2023}%
  \BibitemOpen
  \bibfield  {author} {\bibinfo {author} {\bibfnamefont {C.}~\bibnamefont {González}}, \bibinfo {author} {\bibfnamefont {M.~E.}\ \bibnamefont {Innocenti}}, \ and\ \bibinfo {author} {\bibfnamefont {A.}~\bibnamefont {Tenerani}},\ }\href {\doibase 10.1017/s0022377823000120} {\bibfield  {journal} {\bibinfo  {journal} {Journal of Plasma Physics}\ }\textbf {\bibinfo {volume} {89}} (\bibinfo {year} {2023}),\ 10.1017/s0022377823000120}\BibitemShut {NoStop}%
\bibitem [{\citenamefont {Mininni}\ \emph {et~al.}(2011)\citenamefont {Mininni}, \citenamefont {Rosenberg}, \citenamefont {Reddy},\ and\ \citenamefont {Pouquet}}]{Mininni2011}%
  \BibitemOpen
  \bibfield  {author} {\bibinfo {author} {\bibfnamefont {P.~D.}\ \bibnamefont {Mininni}}, \bibinfo {author} {\bibfnamefont {D.}~\bibnamefont {Rosenberg}}, \bibinfo {author} {\bibfnamefont {R.}~\bibnamefont {Reddy}}, \ and\ \bibinfo {author} {\bibfnamefont {A.}~\bibnamefont {Pouquet}},\ }\href {\doibase https://doi.org/10.1016/j.parco.2011.05.004} {\bibfield  {journal} {\bibinfo  {journal} {Parallel Computing}\ }\textbf {\bibinfo {volume} {37}},\ \bibinfo {pages} {316} (\bibinfo {year} {2011})}\BibitemShut {NoStop}%
\bibitem [{\citenamefont {Mininni}, \citenamefont {Gomez},\ and\ \citenamefont {Mahajan}(2005)}]{Mininni2005}%
  \BibitemOpen
  \bibfield  {author} {\bibinfo {author} {\bibfnamefont {P.~D.}\ \bibnamefont {Mininni}}, \bibinfo {author} {\bibfnamefont {D.~O.}\ \bibnamefont {Gomez}}, \ and\ \bibinfo {author} {\bibfnamefont {S.~M.}\ \bibnamefont {Mahajan}},\ }\href {\doibase 10.1086/426534} {\bibfield  {journal} {\bibinfo  {journal} {The Astrophysical Journal}\ }\textbf {\bibinfo {volume} {619}},\ \bibinfo {pages} {1019–1027} (\bibinfo {year} {2005})}\BibitemShut {NoStop}%
\bibitem [{\citenamefont {Andrés}\ \emph {et~al.}(2018)\citenamefont {Andrés}, \citenamefont {Sahraoui}, \citenamefont {Galtier}, \citenamefont {Hadid}, \citenamefont {Dmitruk},\ and\ \citenamefont {Mininni}}]{Andres2018}%
  \BibitemOpen
  \bibfield  {author} {\bibinfo {author} {\bibfnamefont {N.}~\bibnamefont {Andrés}}, \bibinfo {author} {\bibfnamefont {F.}~\bibnamefont {Sahraoui}}, \bibinfo {author} {\bibfnamefont {S.}~\bibnamefont {Galtier}}, \bibinfo {author} {\bibfnamefont {L.~Z.}\ \bibnamefont {Hadid}}, \bibinfo {author} {\bibfnamefont {P.}~\bibnamefont {Dmitruk}}, \ and\ \bibinfo {author} {\bibfnamefont {P.~D.}\ \bibnamefont {Mininni}},\ }\href {\doibase 10.1017/s0022377818000788} {\bibfield  {journal} {\bibinfo  {journal} {Journal of Plasma Physics}\ }\textbf {\bibinfo {volume} {84}} (\bibinfo {year} {2018}),\ 10.1017/s0022377818000788}\BibitemShut {NoStop}%
\bibitem [{\citenamefont {Orszag}(1971)}]{Orszag1971}%
  \BibitemOpen
  \bibfield  {author} {\bibinfo {author} {\bibfnamefont {S.~A.}\ \bibnamefont {Orszag}},\ }\href {\doibase https://doi.org/10.1002/sapm1971504293} {\bibfield  {journal} {\bibinfo  {journal} {Studies in Applied Mathematics}\ }\textbf {\bibinfo {volume} {50}},\ \bibinfo {pages} {293} (\bibinfo {year} {1971})},\ \Eprint {http://arxiv.org/abs/https://onlinelibrary.wiley.com/doi/pdf/10.1002/sapm1971504293} {https://onlinelibrary.wiley.com/doi/pdf/10.1002/sapm1971504293} \BibitemShut {NoStop}%
\end{thebibliography}%

\end{document}